\documentclass[11pt]{article}
\usepackage{amsfonts}
\usepackage[nosort]{cite}
\csname @addtoreset\endcsname{equation}{section}
\newcommand{\fft}[2]{{\frac{#1}{#2}}}
\newcommand{\ft}[2]{{\textstyle{\frac{#1}{#2}}}}
\newcommand{\sqr}[2]{{\vcenter{\vbox{\hrule height.#2pt
        \hbox{\vrule width.#2pt height#1pt \kern#1pt
        \vrule width.#2pt}\hrule height.#2pt}}}}
\newcommand{\Tr}{\,{\rm Tr}\,}

\newcommand{\be}{\begin{equation}}
\newcommand{\ee}{\end{equation}}
\newcommand{\bea}{\begin{eqnarray}}
\newcommand{\eea}{\end{eqnarray}}
\begin{document}
\begin{titlepage}

\hbox to \hsize{{\tt hep-th/0412043}\hss{\tt MCTP-04-67}}

\vskip 1.7 cm

\begin{center}
{\bf \Large 
Bubbling 1/4 BPS solutions in type IIB  \\[.3cm]
and supergravity reductions on $S^n\times S^n$}

\vskip 1cm

{\bf \large James T. Liu, Diana Vaman and W.Y. Wen}  

\vskip 1cm
{\it Michigan Center for Theoretical Physics,
Randall Laboratory of Physics,\\
The University of Michigan, Ann Arbor, MI 48109--1120}
\vskip .5 cm
{E-mail: \{{\tt jtliu,dvaman,wenw}\}\tt@umich.edu}  

\end{center}

\vspace{1cm}

\begin{abstract}
We extend the construction of bubbling 
1/2 BPS solutions of Lin, Lunin and Maldacena (hep-th/0409174) in two 
directions. 
First we enquire whether
bubbling 1/2 BPS solutions can be constructed in minimal 6d supergravity 
and second we construct solutions that are 1/4 BPS in type IIB.
We find that the $S^1\times S^1$ bosonic reduction of (1,0) 6d supergravity
to 4d gravity coupled to 2 scalars and a gauge field
is consistent only provided that the gauge field obeys a constraint
($F\wedge F=0$). This is to be contrasted to the case of the $S^3\times S^3$
bosonic reduction of type IIB supergravity to 4d gravity, 2 scalars and a
gauge field, where consistency is achieved without imposing any such
constraints. Therefore, in the case of (1,0) 6d supergravity 
we are able to construct 1/2 BPS solutions, similar to those derived in
type IIB, provided that this additional constraint is satisfied. This 
ultimately prohibits the construction of a family of 1/2 BPS solutions
corresponding to a bubbling AdS$_3\times S^3$ geometry.
Returning to type IIB solutions, by turning on an axion-dilaton field we
construct a family of bubbling 1/4 BPS solutions. This corresponds to the
inclusion of back-reacted D7 branes to the solutions of Lin, Lunin and 
Maldacena.  
\end{abstract}

\end{titlepage}

\section{Introduction and Summary}

A recent paper by Lin, Lunin and Maldacena \cite{Lin:2004nb} provided
a nice supergravity 
realization of chiral primary operators in ${\cal N}=4$ super-Yang-Mills. These
operators, with conformal dimension $\Delta$ equal their $U(1)_R$ charge,
form a decoupled sector of BPS states which can be identified
with a gauged quantum mechanics matrix model, with a harmonic 
oscillator potential \cite{Corley:2001zk,Berenstein:2004kk}. 
By going to the
eigenvalue basis, the path integral
measure acquires a Van der Monde determinant factor which makes the eigenvalues
behave as fermions which fill the energy levels inside a harmonic potential
well. There is yet another perspective on the dynamics of eigenvalues: they
correspond to the electrons in a magnetic field which fill the lowest 
Landau levels; by dropping the kinetic term, the positions of the
electrons in the plane become non-commutative/canonical conjugates 
(quantum Hall effect) \cite{Lin:2004nb,Berenstein:2004hw}.

There appears to be a one-to-one correspondence between the phase space regions 
occupied by the eigenvalues, and a similar picture that characterizes the
supergravity solutions. Specifically, the 1/2 BPS family of solutions of
\cite{Lin:2004nb}
is constructed in terms of an auxiliary function; the boundary conditions 
which must be enforced on this auxiliary function in order for the 
supergravity solution to be non-singular reproduce precisely the phase space
configuration of the eigenvalues. On the supergravity side, the 
incompressibility of the ``phase space'' has to be tied to the requirement
that the 5-form RR-flux be fixed. In particular, the 
ground state on the matrix model side is a circular quantum Hall droplet 
in phase space, while on the supergravity side, the same droplet 
corresponds to the AdS$_5\times S^5$ ground state, with the radius of the 
droplet related to the $R_{{\rm AdS}_5}^2$ radius.
Small excitations of the Fermi sea on the matrix side correspond to  
AdS$_5\times S^5$ perturbations by gravitons (ripples on the ground state 
droplet) or giant gravitons (small holes inside the ground state droplet, or 
small droplets outside the ground state droplet) \cite{Lin:2004nb}.

It is worth asking whether a similar picture might carry through for the 
case of AdS$_3\times S^3$ which is the near horizon geometry of a D1-D5
brane configuration, and whether there is a bubbling AdS$_3$ configuration
corresponding to perturbation by giant gravitons. The six-dimensional
giant gravitons are also configurations with $\Delta=J$, but they have 
certain peculiar features: they exist only for a discrete set of values of the 
angular momentum ($J=nN_5+mN_1$, where $N_5, N_1$ are the numbers
of $D_5$, respectively $D_1$ branes), and the potential governing their size 
is flat. The dual gauge theory in this case is a 1+1 dimensional CFT
living on the boundary of AdS$_3$.

To address this question we look for 1/2 BPS solutions to minimal
six-dimensional
supergravity, which have $S^1\times S^1$ isometry. More precisely
we consider the following ansatz:
\begin{eqnarray}
ds^2&=&g_{\mu\nu}(x)dx^\mu dx^\nu+e^{H(x)+G(x)}d\phi^2 + e^{H(x)-G(x)}
d\tilde\phi^2,\nonumber\\
-2 H_{(3)}&=& F_{(2)}\wedge d\phi+\tilde F_{(2)}\wedge d\tilde\phi,
\end{eqnarray}
where $x^\mu=\{t,y,x^1,x^2\}$.  This is a natural extension of the ansatz
used in \cite{Lin:2004nb}, and similarly reduces the problem to that of an
effective four-dimensional theory.
Requiring that this ansatz preserves supersymmetry, 
we found two possible sets of Killing spinors: one which is independent
on the $\phi, \tilde\phi$ coordinates, and which yields a conventional
Kaluza-Klein dimensional reduction, and a second set which carries
Kaluza-Klein momentum
on the two circles. The latter set of Killing spinors gives rise to
six-dimensional solutions which include the AdS$_3\times S^3$ 
and the maximally symmetric plane wave solutions, as well as the multi-center
D1-D5 solutions. The metric and the self-dual 3-form are expressed in terms
of the same auxiliary function as in the 10-dimensional case, namely
$\partial_{x_1}^2 z +\partial_{x_2}^2 z + y \partial_y (1/y \partial_y z)=0$. 
The corresponding metric is non-singular provided that 
the same boundary conditions as in \cite{Lin:2004nb} are imposed
on the auxiliary function: $z(x_1,x_2,y=0)=\pm 1/2$.
However, this time
the field equations are satisfied only if an additional constraint
is enforced as well: $F\wedge F=0$, which can be easily seen by inspecting the 
$\phi\tilde\phi$ component of Einstein equations.
This constraint appears as a non-linear
first order differential equation which the auxiliary
function must satisfy as well: $(\partial_{x_1}z)^2+(\partial_{x_2}z)^2
+(\partial_y z)^2=(1-4z^2)^2/(4y^2)$.
One can check that  for the case of AdS$_3\times S^3$ and that of the 
maximally symmetric plane wave the corresponding 
auxiliary functions do satisfy this additional constraint. However, the 
image of a bubbling AdS$_3$ appears to be incompatible with the additional 
constraint, and we have to conclude that the starting ansatz is too restrictive
to describe giant gravitons on AdS$_3\times S^3$. 
In fact, therein lies the resolution: in order to eliminate this
constraint, we must allow for (at least) an off-diagonal metric component
$g_{\phi\tilde\phi}\neq 0$ in the ansatz. 
This would correspond in the 4-dimensional
reduction language to keeping a non-vanishing axion field besides
the two scalars and the gauge field. We hope to report more on this
in a future work.
We also considered the case when a tensor multiplet (dilaton, dilatino and 
anti-selfdual tensor field) is coupled to six-dimensional minimal supergravity.
With the same metric ansatz as before, we arrived yet again at the same
auxiliary function $z(x^1, x^2,y)$, same metric and the same constraint
$F\wedge F=0$.

Finally, we return to the 10-dimensional type IIB supergravity to investigate
a family of 1/4 BPS solutions which preserve the same set of $SO(4)
\times SO(4)$ isometries as in \cite{Lin:2004nb} and which correspond to
turning on an axion-dilaton field $\tau$, or equivalently,
including the back-reaction of a stack of $N_f$ D7 branes. 
We found it possible to embed
the D7 branes in a way compatible with the metric ansatz of \cite{Lin:2004nb},
provided that they are transverse to $x^1, x^2$. The presence of the D7
branes manifests in the metric by curving the $x^1, x^2$ 
directions with an additional factor $e^{\psi(x^1,x^2)}\propto\Im\tau$.
The auxiliary function
$z(x^1,x^2,y)$ obeys a slightly modified second order differential
equation, but one still imposes the same boundary conditions at $y=0$. 
By setting up a perturbative expansion in $N_f$, we notice that to first order
in $N_f$, and near the D7 branes, their effect on the geometry is to create
a deficit angle in the $x^1, x^2$ plane. Given that the fluctuations in the 
$Z=x^1+i x^2$ direction transverse to the stack of D3 branes 
correspond to the BPS chiral operators
that are singled out in the gauged quantum mechanics matrix model, and 
that their eigenvalues correspond to the electrons in the quantum Hall effect,
we argue that the deficit angle in the supergravity $x^1, x^2$ ``phase space''
can be interpreted as the fractional statistics of the electrons in a 
fractional quantum Hall system. 

It is worth noting that, despite recent advances, the complete
classification of supersymmetric backgrounds with fluxes remains a
technically challenging issue, especially in higher dimensions or
with a large number of supersymmetries.  The analysis of \cite{Lin:2004nb}
is striking in this regards, in that by postulating a minimal
set of isometries, the problem of classification may be greatly simplified.
In particular, by reducing a $(4+2n)$-dimensional system on $S^n\times S^n$,
we end up having to work with only two scalars and a gauge field in
four dimensions.  Of course, it is tempting to view this reduction as
a Kaluza-Klein reduction on $S^n\times S^n$.  In general, care must be taken
when introducing states in the massive Kaluza-Klein tower.
Here, however, by truncating to the singlet sector on $S^n\times S^n$,
one is able to obtain a consistent {\it bosonic} truncation.  This is
sufficient for investigating the supersymmetry of the original system,
provided the full supersymmetry transformations are used.  One unusual
feature of this analysis is the possibility of obtaining bosonic backgrounds
which do not appear supersymmetric in four dimensions, but which are
nevertheless supersymmetric when viewed as a solution of the original
higher dimensional theory.  This phenomenon was noted in \cite{Duff:1997qz},
where it was referred to as `supersymmetry without supersymmetry'.

Furthermore, this method of obtaining gravitational solutions by solving
a harmonic equation in an auxiliary space is similar in spirit to the
work of Weyl \cite{Weyl} (see also \cite{Emparan:2001wk} for a generalization
to higher dimensions), who found all static axisymmetric solutions to
the four-dimensional vacuum Einstein equations by mapping the problem
to a cylindrically symmetric Laplacian problem in an auxiliary flat
three-dimensional space.  What makes the problem tractable in the Weyl
case is the presence of a sufficient number of commuting Killing symmetries.
Of course, one of the interesting features of the Weyl solutions is that
they represent non-supersymmetric configurations of black holes held together
by rods or struts.  Thus, in that case, it is rather surprising that they
may be described by solutions of a harmonic equation.  For BPS configurations
this is somewhat less of a surprise, as they are expected to obey the
principle of linear superposition.  Nevertheless, it is suggestive that
new results in the classification of supersymmetric vacua may be obtained
by revisiting some of the Weyl solution techniques in the present context.
In this sense, it may also be worth looking at the M-theory compactification
of \cite{Lin:2004nb} on $S^2\times S^5$ from the four-dimensional
perspective.

The paper is structured as follows: Section 2 is dedicated to a proof that
the bosonic reduction of type IIB supergravity on $S^3\times S^3$ by retaining
the breathing modes of the two 3-spheres and a gauge field in the reduced 
4-dimensional theory is a consistent {\it bosonic} reduction. We also 
review the supersymmetry analysis of \cite{Lin:2004nb}, using a slightly 
different representation of the 10-dimensional Clifford algebra which allows 
for more streamlined expressions. Section 3, which is organized in the same 
fashion as Section 2, contains the analysis of the six-dimensional supergravity
reduction on $S^1\times S^1$ to the same set of 4-dimensional fields, and
arrives at the conclusion that the reduction is a consistent {\it bosonic}
reduction provided that the constraint $F\wedge F=0$ is satisfied.
In Section 4 we construct the six-dimensional solutions which are 
compatible with supersymmetry. The additional constraint does not allow
for solutions corresponding to bubbling AdS$_3\times S_3$. Section 5 
details the construction of the 1/4 BPS family of solutions 
corresponding to a bubbling 
AdS$_5\times S^5$ in the presence of D7 branes. 

We have included the most 
technical parts of our investigation in a set of appendices.
Appendix A contains a unified treatment of both the type IIB and 
six-dimensional supergravity bosonic reductions on $S^n\times S^n$, where
$n=3(1)$ for the 10(6)-dimensional case respectively. 
Appendix B
discusses the integrability condition of the supersymmetry variations
of both type IIB and six-dimensional supergravity, and highlights the 
difference between the two, in the sense that the 
constraint $F\wedge F=0$ shows up in all $S^n\times S^n$ 
reduction cases other than $n=3$. Finally, Appendix C contains the full set
of differential identities for spinor bilinears implied by supersymmetry.

\section{$S^3\times S^3$ compactification of IIB supergravity}

The bosonic fields of IIB supergravity are given by the NSNS fields
$g_{MN}$, $B_{MN}$ and $\phi$ as well as the RR field strengths $F_{(1)}$,
$F_{(3)}$ and $F_{(5)}^+$.  In the Einstein frame, the IIB action has
the form
\begin{eqnarray}
e^{-1}{\cal L}=R-\fft{\partial_M\tau\partial^M\overline\tau}{2(\Im\tau)^2}
-\fft{G_{(3)}\cdot \overline G_{(3)}}{2\cdot3!\Im\tau}
-\fft{\widetilde F_{(5)}^2}{4\cdot5!}
+\fft1{4i}\fft{C_{(4)}\wedge G_{(3)}\wedge\overline G_{(3)}}{\Im\tau},
\label{eq:iiblag}
\end{eqnarray}
where the self-duality $\widetilde F_{(5)}=*\widetilde F_{(5)}$ must still
be imposed by hand on the equations of motion.  Here the field strengths
are defined by
\begin{eqnarray}
&&G_{(3)}=F_{(3)}-\tau H_{(3)},\nonumber\\
&&F_{(3)}=dC_{(2)},\qquad H_{(3)}=dB_{(2)},\nonumber\\
&&\widetilde F_{(5)}=dC_{(4)}-\ft12C_{(2)}\wedge H_{(3)}+\ft12B_{(2)}
\wedge F_{(3)},
\end{eqnarray}
and $\tau=C_{(0)}+ie^{-\phi}$ is the familiar axion-dilaton.

Although we are not directly concerned with the entire fermionic sector,
since we are interested in the Killing spinor equations, we will need the
IIB gravitino and dilatino variations
\begin{eqnarray}
\delta\Psi_M&=&[D_M+\fft{i}{16\cdot5!}\widetilde F_{NPQRS}\Gamma^{NPQRS}
\Gamma_M]
\varepsilon\nonumber\\
&&\qquad-\fft1{96}(\Gamma_M\Gamma^{NPQ}+2\Gamma^{NPQ}\Gamma_M)G_{NPQ}
\varepsilon^*,\nonumber\\
\delta\lambda&=&i\Gamma^MP_M\varepsilon^*
-\fft{i}{24}G_{MNP}\Gamma^{MNP}\varepsilon.
\label{eq:iibsusy}
\end{eqnarray}
Here $D_M=\nabla_M-\fft{i}2Q_M$ where $P_M$ and $Q_M$ are the scalar kinetic
and composite $U(1)$ connection, respectively.

Note that the NSNS sector of the IIB model can be reduced on $S^3\times
S^3$ \cite{Chamseddine:1997mc,Cvetic:1999au} to yield the gauged $N=4$
Freedman-Schwarz theory \cite{Freedman:1978ra}.  What we are interested
in at present, however, is a reduction with additional degrees of
freedom, in particular the self-dual 5-form as well as metric breathing
modes.  Here we note that, from a Kaluza-Klein point of view, the breathing
modes are actually part of the massive Kaluza-Klein tower.  In general,
in the Freedman-Schwarz supergravity context, massive Kaluza-Klein
supermultiplets necessarily involve charged modes on the spheres.  As a
result, it would be inconsistent to retain a single massive multiplet
without retaining the entire tower.

Nevertheless, it is always possible to obtain a consistent bosonic
breathing mode reduction by retaining only singles on $S^3\times S^3$
\cite{Bremer:1998zp,Liu:2000gk}.  While the truncated theory is
non-supersymmetric, we may still explore the original ten-dimensional
Killing spinor equations obtained from (\ref{eq:iibsusy}), even in
the context of the reduced bosonic fields.  In this fashion, bosonic
solutions of the compactified theory may be lifted to supersymmetric
backgrounds of the original IIB theory, so long as the original Killing
spinor equations are satisfied.

We now follow \cite{Lin:2004nb}, and turn to the sector where only
$\widetilde F_{(5)}$ is turned on (in addition to the metric).  In
this case, the equations of motion obtained from (\ref{eq:iiblag})
admit a consistent truncation, so the relevant ten-dimensional
Lagrangian is of the form
\begin{equation}
e^{-1}{\cal L}=R-\fft1{4\cdot5!}F_{(5)}^2,
\end{equation}
where $F_{(5)}=dC_{(4)}$ and $F_{(5)}=*F_{(5)}$ is to be imposed on
the equations of motion.  This system is now that of a single self-dual
form-field coupled to gravity, admitting a straightforward reduction
on $S^3\times S^3$.  In particular, we take a reduction ansatz
preserving an $SO(4)\times SO(4)$ isometry of the form
\begin{eqnarray}
ds_{10}^2&=&g_{\mu\nu}(x)dx^\mu dx^\nu+e^{H(x)}(e^{G(x)}d\Omega_3^2+e^{-G(x)}
d\widetilde\Omega_3^2),\nonumber\\
F_{(5)}&=&F_{(2)}\wedge\omega_3+\widetilde F_{(2)}\wedge\widetilde\omega_3,
\label{eq:s3s3ans}
\end{eqnarray}
The details of this reduction are given in Appendix~\ref{sec:apa}.  In the
end, we obtain an effective four-dimensional Lagrangian of the form
\begin{equation}
e^{-1}{\cal L}=e^{3H}[R+\ft{15}2\partial H^2-\ft32\partial G^2
-\ft14e^{-3(H+G)}F_{\mu\nu}^2+12e^{-H}\cosh G].
\label{eq:4lag}
\end{equation}

At this point it is worth noting that the model of \cite{Lin:2004nb}
may be extended, not just by turning on the axion-dilaton, but also by
retaining the 3-form field-strength with an ansatz of the form
\begin{equation}
G_{(3)}=G_{(3)}(x)+a(x)\omega_3+\widetilde a(x)\widetilde\omega_3.
\end{equation}
While this may be of interest for obtaining additional supersymmetric
backgrounds, we will not further pursue this direction at present.

\subsection{Supersymmetry variations}

We now turn to the reduction of the IIB supersymmetry variations,
(\ref{eq:iibsusy}).  At present, since we only turn on the metric
and self-dual 5-form, the only non-trivial variation is that of the
IIB gravitino, which has the form
\begin{equation}
\delta\psi_M=[\nabla_M+\ft{i}{16\cdot5!}F_{NPQRS}\Gamma^{NPQRS}\Gamma_M]
\varepsilon.
\end{equation}
Writing this out in components, and using the reduction ansatz
(\ref{eq:s3s3ans}), we find
\begin{eqnarray}
\delta\psi_\mu&=&[\nabla_\mu-\ft1{16}e^{-\fft32(H+G)}F_{\nu\lambda}
\Gamma^{\nu\lambda}\Gamma^{(3)}\Gamma_\mu]\varepsilon,\nonumber\\
\delta\psi_a&=&[\hat\nabla_a+\ft14\Gamma_a\Gamma^\mu\partial_\mu(H+G)
-\ft1{16}e^{-\fft32(H+G)}F_{\mu\nu}\Gamma^{\mu\nu}\Gamma^{(3)}\Gamma_a]
\varepsilon,\nonumber\\
\delta\psi_{\tilde a}&=&[\hat\nabla_{\tilde a}+\ft14\Gamma_{\tilde a}
\Gamma^\mu\partial_\mu(H-G)-\ft1{16}e^{-\fft32(H+G)}F_{\mu\nu}
\Gamma^{\mu\nu}\Gamma^{(3)}\Gamma_{\tilde a}]\varepsilon,
\end{eqnarray}
where $\Gamma^{(3)}=-\fft{i}6\epsilon_{abc}\Gamma^{abc}$, and we have
taken into account the chirality of IIB spinors, $\Gamma^{11}\varepsilon
=\varepsilon$, where $\Gamma^{11}=\fft1{10!}\epsilon_{M_1\cdots M_{10}}
\Gamma^{M_1\cdots M_{10}}$.

To proceed, we choose a Dirac decomposition respecting the $S^3\times S^3$
symmetry
\begin{eqnarray}
\Gamma_\mu&=&\gamma_\mu\times1\times1\times\sigma_1,\nonumber\\
\Gamma_a&=&1\times\sigma_a\times1\times\sigma_2,\nonumber\\
\Gamma_{\tilde a}&=&\gamma_5\times1\times\sigma_{\tilde a}\times\sigma_1.
\end{eqnarray}
Here the $\sigma$'s are ordinary Pauli matrices, while $\gamma_5=\fft{i}{4!}
\epsilon_{\mu\nu\rho\sigma}\gamma^{\mu\nu\rho\sigma}$.  It is straightforward
to see that, in this representation, the respective `chirality' matrices are
\begin{eqnarray}
\Gamma^{(5)}&=&\fft{i}{4!}\epsilon_{\mu\nu\rho\sigma}\Gamma^{\mu\nu\rho\sigma}
=\gamma_5\times1\times1\times1,\nonumber\\
\Gamma^{(3)}&=&-\fft{i}{3!}\epsilon_{abc}\Gamma^{abc}=1\times1\times1\times
\sigma_2,\nonumber\\
\Gamma^{(\tilde 3)}&=&-\fft{i}{3!}\epsilon_{\tilde a\tilde b\tilde c}
\Gamma^{\tilde a\tilde b\tilde c}=\gamma_5\times1\times1\times\sigma_1,
\nonumber\\
\Gamma^{11}&=&\fft1{10!}\epsilon_{M_1\cdots M_{10}}\Gamma^{M_1\cdots M_{10}}
=i\Gamma^{(5)}\Gamma^{(3)}\Gamma^{(\tilde 3)}=1\times1\times1\times\sigma_3.
\end{eqnarray}

With this decomposition, the gravitino transformation becomes
\begin{eqnarray}
\delta\psi_\mu&=&[\nabla_\mu+\ft{i}{16}e^{-\fft32(H+G)}F_{\nu\lambda}
\gamma^{\nu\lambda}\gamma_\mu]\epsilon,\nonumber\\
\delta\psi_a&=&\hat\nabla_a\epsilon-\fft{i}4\sigma_a[\gamma^\mu\partial_\mu(H+G)
-\ft{i}4e^{-\fft32(H+G)}F_{\mu\nu}\gamma^{\mu\nu}]\epsilon,\nonumber\\
\delta\psi_{\tilde a}&=&\hat\nabla_{\tilde a}\epsilon+\fft14\gamma_5
\sigma_{\tilde a}[\gamma^\mu\partial_\mu(H-G)+\ft{i}4e^{-\fft32(H+G)}F_{\mu\nu}
\gamma^{\mu\nu}]\epsilon.
\label{eq:psiform}
\end{eqnarray}
We now write the complex IIB spinor as $\varepsilon=\epsilon\times\chi
\times\widetilde\chi$ where $\chi$ and $\widetilde\chi$ are Killing spinors
on the respective unit three-spheres.  Since they satisfy the Killing
spinor equations
\begin{equation}
[\hat\nabla_a+\fft{i\eta}2\hat\sigma_a]\chi=0,\qquad
[\hat\nabla_{\tilde a}+\fft{i\widetilde\eta}2
\hat\sigma_{\tilde a}]\widetilde\chi=0,
\label{eq:ksesphere}
\end{equation}
where $\eta=\pm1$ and $\widetilde\eta=\pm1$, the transformations
(\ref{eq:psiform}) may be rewritten as
\begin{eqnarray}
\delta\psi_\mu&=&[\nabla_\mu+\ft{i}{16}e^{-\fft32(H+G)}F_{\nu\lambda}
\gamma^{\nu\lambda}\gamma_\mu]\epsilon,\label{psimu}\\
\delta\chi_H&=&[\gamma^\mu\partial_\mu H+e^{-\fft12H}(\eta e^{-\fft12G}
-i\widetilde\eta\gamma_5e^{\fft12G})]\epsilon,\label{ch}\\
\delta\chi_G&=&[\gamma^\mu\partial_\mu G-\ft{i}4e^{-\fft32(H+G)}F_{\mu\nu}
\gamma^{\mu\nu}+e^{-\fft12H}(\eta e^{-\fft12G}+i\widetilde\eta\gamma_5
e^{\fft12G})]\epsilon.\label{cg}
\end{eqnarray}
Here we have defined
\begin{eqnarray}
\psi_a&=&-\fft{i}4\sigma_a(\chi_H+\chi_G),\nonumber\\
\psi_{\tilde a}&=&\fft14\gamma_5\sigma_{\tilde a}(\chi_H-\chi_G).
\end{eqnarray}

To summarize, we have achieved a consistent bosonic breathing-mode
reduction of the truncated IIB theory on $S^3\times S^3$.  The resulting
four-dimensional Lagrangian is given by (\ref{eq:4lag}), while the
reduction of the supersymmetry variations results in the system
(\ref{psimu})--(\ref{cg}).  Note that, while the variations have
a typical form associated with four-dimensional ${\cal N}=2$ supergravity,
they nevertheless should not to be thought of as supersymmetries of the
effective four-dimensional theory.  This is because the bosonic
truncation to singlets on $S^3\times S^3$ does not (and cannot) retain
the complete supermultiplet content in the massive Kaluza-Klein sector,
as indicated previously.

By reducing to an effective four-dimensional theory, we have ended
up with a fairly simple system to investigate.  Of course, we
are mainly concerned with solving the Killing spinor equations derived
from (\ref{psimu})--(\ref{cg}).  In contrast to the 
approach of \cite{Lin:2004nb},
we may now work directly in a four-dimensional context, even though the
equations originated from the full IIB gravitino variation,
(\ref{eq:iibsusy}).  In addition, by reducing the six-dimensional ${\cal N}
=(1,0)$ solution to four dimensions, we would similarly obtain an effective
theory of the same general form as (\ref{eq:4lag}) and
(\ref{psimu})--(\ref{cg}).
Thus, using the four-dimensional picture, we will be able to solve both
model simultaneously.

\section{$S^1\times S^1$ compactification of Minimal $D=6$ Supergravity}
\label{sec:s1s1}

We now turn to the case of minimal $D=6$, ${\cal N}=(1,0)$ supergravity
which admits an AdS$_3\times S^3$ solution corresponding to the near
horizon of the D1-D5 system.  The field content of this theory is
given by the gravity multiplet ($g_{MN}$, $\psi_M$, $B_{MN}^+$), where
$B_{MN}^+$ denotes a two-form potential with self-dual field strength,
$H_{(3)}=*H_{(3)}$ where $H_{(3)}=dB_{(2)}^+$.
This is often extended by the addition of a dilaton multiplet
($B_{MN}^-$, $\lambda$, $\phi$), as the inclusion of both chiralities of
$B_{MN}$ then allow a covariant Lagrangian formulation.  Furthermore,
the Salam-Sezgin model \cite{Salam:1984cj} may be obtained by coupling to
an Abelian vector multiplet ($A_\mu$, $\chi$).  In the following section,
we will consider the addition of the dilaton multiplet.  Here, however,
we only focus on the minimal supergravity without dilaton or vector
multiplet.

The bosonic sector of the minimal theory may be described by the Lagrangian
\begin{equation}
e^{-1}{\cal L}=R-\fft1{2\cdot3!}H_{(3)}^2
\end{equation}
where the self-duality condition on $H_{(3)}$ remains to be imposed
after deriving the equations of motion.  Note that, for convenience,
we have chosen to normalize $H_{(3)}$ as if it were an unconstrained
form-field.  The resulting equations are simply
\begin{equation}
R_{MN}=\ft14H_{MPQ}H_N{}^{PQ},\qquad H_{(3)}=*H_{(3)},\qquad dH_{(3)}=0.
\end{equation}
In addition, we note that the gravitino variation is given by
\begin{equation}
\delta\psi_M=[\nabla_M+\ft1{48}H_{NPQ}\Gamma^{NPQ}\Gamma_M]\varepsilon.
\label{eq:6gto}
\end{equation}
These are the starting points of the reduction.

Analogous to the $S^3\times S^3$ reduction of IIB supergravity, we
reduce the ${\cal N}=(1,0)$ theory on $S^1\times S^1$.  This is, of
course, a familiar situation, as it is simply an ordinary Kaluza-Klein
reduction on $T^2$, specialized to a rectangular torus.  Before
proceeding, it is worthwhile recalling the standard Kaluza-Klein result.
Since this supergravity involves eight real supercharges, it reduces to
a ${\cal N}=2$ theory in four dimensions.  Since the six-dimensional
metric reduces to a four-dimensional metric, two vectors and three
scalars (two dilatonic and one axionic), and the self-dual $B_{MN}^+$
reduces to a vector and axionic scalar, the resulting four-dimensional
theory consists of ${\cal N}=2$ supergravity coupled to two vector
multiplets.

In the present case, however, we specify a reduction ansatz of the form
\begin{eqnarray}
ds_{6}^2&=&g_{\mu\nu}(x)dx^\mu dx^\nu+e^{H(x)}(e^{G(x)}d\phi^2+e^{-G(x)}
d\widetilde\phi^2),\nonumber\\
H_{(3)}&=&-\ft12F_{(2)}\wedge d\phi-\ft12\widetilde F_{(2)}
\wedge d\widetilde\phi,
\label{eq:s1s1ans}
\end{eqnarray}
which preserves the $SO(2)\times SO(2)$ isometry.  The $-1/2$ factors
in the $H_{(3)}$ ansatz are physically unimportant, but are chosen for
later notational convenience in the reduced supersymmetry variations.
In the framework of
a torus reduction, this ansatz corresponds to setting both metric gauge
fields and both axionic scalars to zero.  As a result, this ends up
corresponding to an inconsistent truncation of the original ${\cal N}=
(1,0)$ theory.  Nevertheless, as we demonstrate below, this inconsistency
manifests itself solely in one additional constraint that must be
imposed by hand on the effective four-dimensional theory.

Given the reduction ansatz (\ref{eq:s1s1ans}), it is straightforward
to apply the results of Appendix~\ref{sec:apa} (while taking into account
the normalization of $F_{(2)}$) to arrive at the effective
four-dimensional Lagrangian
\begin{equation}
e^{-1}{\cal L}=e^H[R+\ft12\partial H^2-\ft12\partial G^2-\ft18e^{-(H+G)}
F_{\mu\nu}^2],
\label{eq:6to4lag}
\end{equation}
where we note that $\widetilde F_{(2)}=-*_4e^{-G}F_{(2)}$.  If desired,
$F_{(2)}$ may be canonically normalized by taking $F_{(2)}\to\sqrt{2}
F_{(2)}$.  However, it will be clear below when considering the
supersymmetry variations why we avoid this last step.
In addition, the reduction of the $R_{\phi\tilde\phi}$ component of the
Einstein equation results in a constraint $F_{(2)}\wedge F_{(2)}=0$.
This is the
manifestation of the inconsistency in the reduction alluded to above.
Ordinarily $F_{(2)}\wedge F_{(2)}$ will source one of the axions.  However, by
truncating them away, we can no longer allow such a source.  Nevertheless,
so long as we satisfy this constraint, all solutions to the effective
four-dimensional theory may be lifted to provide solutions to the original
${\cal N}=(1,0)$ model.

This reduction differs from the $S^3\times S^3$ case since, unlike the
spheres, the circles are flat.  Thus no potential is generated in the
reduced theory.  For the same reason, the scalar $H$, which plays the
r\^ole of a breathing mode in the sphere reduction, is instead an
ordinary massless scalar in the present case.

\subsection{Supersymmetry variations}

Having completed the reduction of the bosonic sector, we now proceed to
reduce the gravitino variation, (\ref{eq:6gto}).  Although the ${\cal N}
=(1,0)$ theory is generally formulated with symplectic-Majorana Weyl
spinors, here the six-dimensional spinors may be taken to be simply
complex Weyl, satisfying the left-handed projection $\Gamma^7\varepsilon=
-\varepsilon$ as well as $\Gamma^7\psi_M=-\psi_M$, where
$\Gamma^7=\fft1{6!}\epsilon_{M_1\cdots M_{6}}\Gamma^{M_1\cdots M_{6}}$.

To obtain an effective four-dimensional description of the supersymmetry,
we find it convenient to decompose the six-dimensional Dirac matrices
according to
\begin{eqnarray}
\Gamma_\mu&=&\gamma_\mu\times\sigma_1,\nonumber\\
\Gamma_4&=&1\times\sigma_2,\nonumber\\
\Gamma_5&=&\gamma_5\times\sigma_1.
\end{eqnarray}
Indices $4$ and $5$ correspond to coordinates $\phi$ and $\widetilde\phi$,  respectively.  It is straightforward to see that
\begin{eqnarray}
\Gamma^{(5)}&=&\fft{i}{4!}\epsilon_{\mu\nu\rho\sigma}
\Gamma^{\mu\nu\rho\sigma}=\gamma_5\times 1 ,\nonumber\\
\Gamma^7&=&\fft1{6!}\epsilon_{M_1\cdots M_{6}}\Gamma^{M_1\cdots M_{6}}
=-1\times\sigma_3.
\end{eqnarray}
As a result, left-handed six-dimensional spinors may be written as
$\varepsilon=\epsilon\times \left[1\atop0\right]$.

Noting that
\begin{equation}
H_{MNP}\Gamma^{MNP}=-\ft32e^{-\fft12(H+G)}F_{\mu\nu}\Gamma^{\mu\nu}
\Gamma^{\underline4}(1+\Gamma^7),
\end{equation}
[taking into account the unusual normalization of (\ref{eq:s1s1ans})]
and using the above Dirac decomposition, the gravitino variation becomes
\begin{eqnarray}
\delta\psi_{\mu}&=&[\nabla_{\mu}+\ft{i}{16}e^{\fft12(H+G)}F_{\nu\lambda}
\gamma^{\nu\lambda}\gamma_{\mu}]\epsilon,\nonumber\\
\delta\psi_\phi&=&[\partial_\phi-\ft{i}{4}e^{\fft12(H+G)}\gamma^{\mu}
\partial_{\mu}(H+G)-\ft{1}{16}F_{\mu\nu}\gamma^{\mu\nu}]\epsilon,\nonumber\\
\delta\psi_{\tilde\phi}&=&[\partial_{\tilde\phi}
+\ft{1}{4}e^{\ft{1}{2}(H-G)}\gamma^5\gamma^{\mu}
\partial_{\mu}(H-G)+\ft{i}{16}e^{-G}F_{\mu\nu}\gamma^5\gamma^{\mu\nu}]\epsilon
\label{ssusy}
\end{eqnarray}
Up to this point, we have followed a conventional Kaluza-Klein reduction
which involves truncation to zero-modes only on $S^1\times S^1$.  However,
we are not necessarily interested in obtaining a consistent truncation
to four dimensions, but rather wish to obtain supersymmetric configurations
of the original ${\cal N}=(1,0)$ theory.  We thus relax the Kaluza-Klein
condition on the supersymmetry parameter $\epsilon$.  In particular, we take
$\epsilon(x,\phi,\tilde\phi)=e^{i(q\phi+\tilde q\tilde\phi)}\epsilon(x)$,
where the Kaluza-Klein momenta $q$ and $\tilde q$ are quantized in
half-integer units (as appropriate for a spinor on a circle).

For convenience of notation, we redefine the (half-integer) Kaluza-Klein
charges as $q=-\eta/2$ and $\tilde q=-\tilde\eta/2$.  Then, for spinors
charged on the two circles, we may make a simple replacement
\begin{equation}
\partial_\phi\to -i\ft{\eta}{2},\qquad\partial_{\tilde\phi}\to
-i\ft{\tilde\eta}{2}.
\end{equation}
As a result, the above transformations may be rewritten as
\begin{eqnarray}
\delta\psi_{\mu}&=&[\nabla_{\mu}+\ft{i}{16}e^{-\ft{1}{2}(H+G)}F_{\nu\lambda}\gamma^{\nu\lambda}\gamma_{\mu}]\epsilon,
\nonumber\\
\delta\chi_H&=&[\gamma^{\mu}\partial_{\mu}H+e^{-\fft12H}
(\eta e^{-\fft12G}-i\tilde\eta\gamma_5 e^{\fft12G})]\epsilon,\nonumber\\
\delta\chi_G&=&[\gamma^{\mu}\partial_{\mu}G-\ft{i}4e^{-\fft12(H+G)}F_{\mu\nu}\gamma^{\mu\nu}+e^{-\fft12H}
(\eta e^{-\fft12G}+i\tilde\eta\gamma_5 e^{\fft12G})]\epsilon,
\label{eq:6to4susy}
\end{eqnarray}
where we have defined the linear combinations
\begin{eqnarray}
\psi_\phi&=&-\ft{i}4e^{\fft12(H+G)}(\chi_H+\chi_G),\nonumber\\
\psi_{\tilde\phi}&=&\ft14e^{\fft12(H-G)}\gamma^5(\chi_H-\chi_G).
\end{eqnarray}
In this form, the supersymmetry variations (\ref{eq:6to4susy}) resemble
those of the $S^3\times S^3$ reduced IIB theory,
(\ref{psimu})--(\ref{cg}).
However, in this case, the parameters $\eta$ and $\tilde\eta$ may take on
any integer values include zero.  Ordinary Killing spinors of the
massless sector Kaluza-Klein reduction are obtained for $\eta=\tilde\eta=0$,
while charged Killing spinors are obtained otherwise.  We keep in mind,
however, that the bosonic sector is unchanged and always has the form of
(\ref{eq:6to4lag}) regardless of the structure of the Killing spinors.

This reduction framework provides yet another example where the 
supersymmetry of spaces such as AdS$_3\times S^3$, viewed as a
fibration, involve Killing spinors charged along $U(1)$ fibers.  Proper
supersymmetry counting then involves proper identification of the
fiber $U(1)$ charges \cite{Duff:1997qz,Duff:1998cr,Liu:2000gk}.

\section{Supersymmetry analysis}

In the previous two sections, we have demonstrated that both the reduction
of IIB theory on $S^3\times S^3$ and the reduction of ${\cal N}=(1,0)$
supergravity on $S^1\times S^1$ lead to similar effective four-dimensional
descriptions.  In particular, this similarity is evident not only in the
bosonic sectors (\ref{eq:4lag}) and (\ref{eq:6to4lag}) but also in the
supersymmetry transformations (\ref{psimu})--(\ref{cg}) and
(\ref{eq:6to4susy}).  At first sight,
this is actually somewhat surprising, as the details of the fermionic
sector, and in particular the mechanics of the supersymmetry algebra, are 
very much dimension dependent.  However, given that both theories truncate
to an identical field content, it is perhaps not unreasonable to expect
that the effective four-dimensional descriptions of the supersymmetry
transformations necessarily have a similar form.

In fact, comparing (\ref{psimu})--(\ref{cg}) with (\ref{eq:6to4susy}),
we find that they may both be written in the form
\begin{eqnarray}
\delta\psi_\mu&=&[\nabla_\mu+\ft{i}{16}e^{-\fft{n}2(H+G)}F_{\nu\lambda}
\gamma^{\nu\lambda}\gamma_\mu]\epsilon\nonumber\\
\delta\chi_H&=&[\gamma^\mu\partial_\mu H+e^{-\fft12H}(\eta e^{-\fft12G}
-i\widetilde\eta\gamma_5e^{\fft12G})]\epsilon,\nonumber\\
\delta\chi_G&=&[\gamma^\mu\partial_\mu G-\ft{i}4e^{-\fft{n}2(H+G)}F_{\mu\nu}
\gamma^{\mu\nu}+e^{-\fft12H}(\eta e^{-\fft12G}+i\widetilde\eta\gamma_5
e^{\fft12G})]\epsilon,
\label{eq:4kse}
\end{eqnarray}
where $n=3$ or $1$ corresponds to the $S^3\times S^3$ or $S^1\times S^1$
cases, respectively.  This description of supersymmetry allows for a
unified analysis of half-BPS solutions, using the methods of
\cite{Lin:2004nb,Gauntlett:2002sc,Gauntlett:2002nw,GMR,Gauntlett:2004zh}%
\footnote{Note that general solutions of minimal ${\cal N}=(1,0)$
supergravity have been classified in \cite{GMR}.  Solutions to IIB theory
on $S^3\times S^3$ were of course examined in \cite{Lin:2004nb}.}.
There are several differences to note about the two cases, however.
Firstly, for $n=3$, the choice of $\eta=\pm1$ and $\widetilde\eta=\pm1$
is required based on satisfying the Killing spinor equations on the
two $3$-spheres.  However, for $n=1$, the parameters $\eta$ and $\widetilde
\eta$ refer to $U(1)$ charges along the two circles, and may be chosen
to be arbitrary integers including zero (to be consistent with charge
quantization).  Secondly, although $F_{(2)}$ shows up identically in
(\ref{eq:4kse}), there is actually a factor of two difference in the
field strength terms in the bosonic Lagrangians, (\ref{eq:4lag}) and
(\ref{eq:6to4lag}).  This factor conspires with the $n$-dependence in
the exponential prefactors $e^{-\fft{n}2(H+G)}$ in (\ref{eq:4kse}) so
as to yield the appropriate (distinct) bosonic equations of motion based
on integrability of the supersymmetry variations.  This issue of
integrability is investigated in Appendix~\ref{sec:apb}.  Finally,
while not directly evident above, it is important to realize that when
$n=1$ there is an added constraint, $F_{(2)}\wedge F_{(2)}=0$, that must
be checked for the solution.

Regardless of the differences mentioned above, we begin with a unified
treatment of both cases.  Our analysis parallels that of \cite{Lin:2004nb}.
Thus we first assume that $\epsilon$ is a Killing
spinor and proceed by forming the spinor bilinears
\begin{eqnarray}
&&f_1=\bar\epsilon\gamma^5\epsilon,\qquad f_2=i\bar\epsilon\epsilon,
\nonumber\\
&&K^\mu=\bar\epsilon\gamma^\mu\epsilon,\qquad L^\mu=\bar\epsilon\gamma^\mu
\gamma^5\epsilon,\nonumber\\
&&Y_{\mu\nu}=i\bar\epsilon\gamma_{\mu\nu}\gamma^5\epsilon.
\label{eq:bilinear}
\end{eqnarray}
The factors of $i$ are chosen to make these tensor quantities real.
Then, by Fierz rearrangement, we may prove the algebraic identities
\begin{equation}
L^2=-K^2=f_1^2+f_2^2,\qquad K\cdot L=0.
\label{eq:h2}
\end{equation}

Next we turn to the differential identities.  While the complete set
of such identities are provided in Appendix~\ref{sec:apc}, only a
subset suffices for the present analysis.  Without yet making any
assumptions about the metric, we may first fix the form of the scalar
quantities $f_1$ and $f_2$.  Combining the differential identities
for $\nabla_\mu f_1$ and $\nabla_\mu f_2$ in (\ref{eq:difdi}) with
the $L_\mu$ identities in (\ref{eq:difhi}) and (\ref{eq:difgi}), we
obtain
\begin{eqnarray}
\partial_\mu f_1&=&\ft14e^{-\fft{n}2(H+G)}*F_{\mu\nu}K^\nu
=\ft12f_1\partial_\mu(H-G),\nonumber\\
\partial_\mu f_2&=&-\ft14e^{-\fft{n}2(H+G)}F_{\mu\nu}K^\nu
=\ft12f_2\partial_\mu(H+G),
\label{eq:fkeqn}
\end{eqnarray}
which may be integrated to obtain
\begin{equation}
f_1=be^{\fft12(H-G)},\qquad f_2=ae^{\fft12(H+G)}.
\end{equation}
The constants $a$ and $b$ are related through the identity $\eta f_2
=-\widetilde\eta e^Gf_1$ of (\ref{eq:difhi}).  In particular
\begin{equation}
a\eta+b\widetilde\eta=0.
\label{eq:abeta}
\end{equation}
Note that at this stage we still allow $\eta$ and $\widetilde\eta$ to be
arbitrary integers (including zero).

Having fixed the scalars $f_1$ and $f_2$, we now turn to the vectors $K_\mu$
and $L_\mu$.  Here, we observe from (\ref{eq:difdi}) that the equations
for $K_\mu$ and $L_\mu$ indicate that $K_{(\mu;\nu)}=0$ (so that $K^\mu$
is a Killing vector) and $dL=0$.  As done in \cite{Lin:2004nb},
this allows us to specialize the metric ansatz below.  Before doing so,
however, we note that these vectors are necessarily normalized by
(\ref{eq:h2}) to be
\begin{equation}
L^2=-K^2=f_1^2+f_2^2=e^H(a^2e^G+b^2e^{-G}).
\label{eq:knorm}
\end{equation}
In particular, this indicates that $K^\mu$ is in fact a
time-like Killing vector%
\footnote{It is useful to observe that this is in agreement with \cite{GMR}
where the 6-dimensional Killing vector $\bar\varepsilon \Gamma^M\varepsilon$, 
$M=0,\dots 5$ was shown to be null. The latter statement can be rephrased
using our representation of 6-dimensional Dirac matrices as
$(\bar\epsilon\gamma^\mu\epsilon)^2+f_1^2+f_2^2=0$.}.
Furthermore, the $L_\mu$ equations of (\ref{eq:difhi}) provide the constraints
\begin{equation}
\eta L_\mu=b\partial_\mu e^H,\qquad\widetilde\eta L_\mu=-a\partial_\mu e^H.
\label{eq:diffh}
\end{equation}

Following \cite{Lin:2004nb}, we now choose a preferred coordinate basis
so that the Killing vector $K^\mu\partial_\mu$ corresponds to
$\partial/\partial t$ and the closed one-form $L_\mu dx^\mu$ to $dy$,
where $t$ and $y$ are two of the four
coordinates.  In particular, we write down the four-dimensional metric as
\begin{equation}
ds_{4}^2=-h^{-2}(dt+V_idx^i)^2+h^2(dy^2+\delta_{ij}dx^idx^j),
\label{eq:ds4}
\end{equation}
where $i,j=1,2$.  Note that we have already taken the spatial metric to
be conformally flat
based on the identical reasoning as \cite{Lin:2004nb}.  The remaining
components of the metric are $V_i$, to be determined below, and $h^2$,
given from (\ref{eq:knorm}) to be
\begin{equation}
h^{-2}=e^H(a^2e^G+b^2e^{-G}).
\label{eq:hexp}
\end{equation}
In addition, for $L=dy$, (\ref{eq:diffh}) yields the constraints
\begin{equation}
\eta=b\partial_ye^H,\qquad\widetilde\eta=-a\partial_ye^H.
\label{eq:ehcon}
\end{equation}
where we still allow for any of these constants $\eta$, $\widetilde\eta$,
$a$ or $b$ to be zero.

At this stage, we have sufficient information to fix the form of the
field strengths $F_{(2)}$ as well as $dV$.  For $F_{(2)}$, we use the
component relations
\begin{equation}
F_{\mu\nu}K^\nu=-\fft{4a}{n+1}\partial_\mu e^{\fft{n+1}2(H+G)},\qquad
\widetilde F_{\mu\nu}K^\nu=-\fft{4b}{n+1}\partial_\mu e^{\fft{n+1}2(H-G)},
\end{equation}
obtained from (\ref{eq:fkeqn}) as well as the explicit form of the
metric (\ref{eq:ds4}) to find
\begin{eqnarray}
F_{(2)}&=&-\fft{4a}{n+1}d\left(e^{\fft{n+1}2(H+G)}\right)\wedge(dt+V)
-\fft{4b}{n+1}h^2e^{n G}*_3d\left(e^{\fft{n+1}2(H-G)}\right),\nonumber\\
\widetilde F_{(2)}&=&-\fft{4b}{n+1}d\left(e^{\fft{n+1}2(H-G)}\right)
\wedge(dt+V)
+\fft{4a}{n+1}h^2e^{-n G}*_3d\left(e^{\fft{n+1}2(H+G)}\right),
\label{eq:f2form}
\end{eqnarray}
where $*_3$ denotes the Hodge dual with respect to the flat spatial metric.
For $dV$, we take the antisymmetric part of $\nabla_\mu K_\nu$ in
(\ref{eq:difdi}), written in form notation as
\begin{equation}
dK=\ft12e^{-\fft{n}2(H+G)}(f_2F_{(2)}-f_1*F_{(2)}),
\end{equation}
and substitute in the expressions for the Killing vector
$K=-h^{-2}(dt+V)$ as well as for $F_{(2)}$.
This gives both the known expression for $h^{-2}$, namely
(\ref{eq:hexp}), as well as the relation
\begin{equation}
dV=-2abh^4e^H*_3dG.
\label{eq:dv}
\end{equation}

\subsection{Specialization of $\eta$ and $\widetilde\eta$}

Until now, we have allowed for all possible values of $\eta$ and
$\widetilde\eta$.  For sphere, as opposed to circle, compactifications
($n>1$) the only possible choices of $\eta$ and $\widetilde\eta$ are
$\pm1$, as this is dictated by the Killing spinor equations on the
sphere, (\ref{eq:ksesphere}).  On the other hand, when $n=1$, we may
consider three distinct possibilities: both non-vanishing, one
vanishing, and both vanishing.

\subsubsection{Both $\eta$ and $\widetilde\eta$ non-vanishing}

We begin with the case of
both $\eta$ and $\widetilde\eta$ non-vanishing.  In particular, to satisfy
(\ref{eq:abeta}), we take $a=-\widetilde\eta$ and $b=\eta$ with $a^2=b^2=1$.
The actual choice of $a =\pm b$ corresponds to the case of chiral
primaries with $\Delta \mp J=0$ in the dual field theory.  For this case,
(\ref{eq:ehcon}) yields the simple result $e^H=y$, so that (\ref{eq:hexp})
becomes $h^{-2}=2y\cosh G$ \cite{Lin:2004nb}.  As in \cite{Lin:2004nb}, the
consistency condition $d^2V=0$ following from (\ref{eq:dv}) yields the
second order differential equation $d(y^{-1}*_3dz)=0$ or
\begin{equation}
(\partial_i^2+y\partial_y\ft1y\partial_y)z=0,
\label{eq:harmz}
\end{equation}
where $z=\fft12\tanh G$.

Of course, it is not surprising that the analysis of \cite{Lin:2004nb}
may be generalized away from $n=3$, as the Killing spinor equations
(\ref{eq:4kse}) have a relatively straightforward dependence on $n$.
However, it is important to examine the complete consistency of the
solution generated above, as in general solving the Killing spinor
equations does not automatically yield a complete solution to the equations
of motion, but only guarantees that a subset of the equations are solved.
This issue of integrability is examined in detail in Appendix~\ref{sec:apb}.
Here, it is sufficient to note that the $F_{(2)}$ equations of
motion are not obviously satisfied.  Instead, by combining (\ref{eq:f2form})
with (\ref{eq:dv}), we find that
\begin{equation}
dF_{(2)}=(n-3)bh^2ye^{\fft{n-1}2(H+G)}(dG\wedge*_3dG-dH\wedge*_3dH).
\end{equation}
This demonstrates that, at least for $n\ne3$, we must impose the additional
constraint $dG\wedge*_3dG=dH\wedge*_3dH$, or
\begin{equation}
\partial_iz\partial_iz+\partial_yz\partial_yz=\fft{(1-4z^2)^2}{4y^2}.
\label{eq:bad}
\end{equation}
That this constraint shows up for $n\ne3$ is related to the fact that
the bosonic equations pick up a $F_{(2)}\wedge F_{(2)}=0$ constraint
in this case as well.
Unfortunately, this non-linear constraint is highly restrictive for
functions $z(x_1,x_2,y)$ already satisfying the Laplacian equation
of motion (\ref{eq:harmz}).  While we have not made an exhaustive search,
we have only found the maximally symmetric AdS$_3\times S^3$ and plane-wave
solutions to satisfy this constraint.

Without loss of generality, we choose $a=b=1$ for 1/2 BPS
solutions.  In this case, the generalization of \cite{Lin:2004nb} for
$n=1$ as well as $3$ may be summarized as follows:
\begin{eqnarray}
ds_6^2 &=& -h^{-2}(dt+V_idx^i)^2
+h^2(dy^2+\delta_{ij}dx^idx^j)+y(e^Gd\Omega_n^2+e^{-G}d\tilde\Omega_n^2),\\
F_{(2)} &=&-\fft4{n+1}\left[
d\left(y^{\fft{n+1}2}e^{\fft{n+1}2G}\right)\wedge (dt+V)
+h^2e^{nG}*_3d\left(y^{\fft{n+1}2}e^{-\fft{n+1}2G}\right)\right],
\end{eqnarray}
where
\begin{equation}
h^{-2} = 2y\cosh{G},\qquad z\equiv \ft12 \tanh{G},\qquad
dV = -\ft{1}{y}*_3dz.
\end{equation}
Note that $F_{(2)}$ is only canonically normalized for $n=3$.  Furthermore,
the function $z$ must satisfy (\ref{eq:harmz}) for any $n$, and additionally
the constraint (\ref{eq:bad}) for $n\ne3$.

\subsubsection{Only one of $\eta$ and $\widetilde\eta$ non-vanishing}

This and the subsequent possibility only applies when $n=1$.  Without loss
of generality, we take $\eta=0$, $\widetilde\eta=\pm1$.  In this case, the
constraint (\ref{eq:abeta}) indicates that $b=0$, so that in particular
$f_1=0$ or $\bar\epsilon\gamma^5\epsilon=0$.  To avoid the degenerate
situation, we assume that $a\ne0$.  Taking $a=-\widetilde\eta$, we see
that (\ref{eq:ehcon}) again gives $e^H=y$.  This time, however, the relation
(\ref{eq:hexp}) yields a single exponential, $h^{-2}=ye^G$.  In addition,
the field strength $F_{(2)}$ is then given by (\ref{eq:f2form})
\begin{equation}
F_{(2)}=-2a\,d(ye^G)\wedge(dt+V),
\end{equation}
indicating that it is of pure electric type.

For both $\eta$ and $\widetilde\eta$ non-vanishing, the second order
equation giving the bubbling picture was obtained from the consistency
condition $d^2V=0$.  Here, however, $dV$ is trivially closed, as may
be seen by setting $b=0$ in (\ref{eq:dv}).  Nevertheless, we must still
satisfy the equation of motion for $F_{(2)}$, most conveniently expressed
as $d\widetilde F_{(2)}=0$ where
\begin{equation}
\widetilde F_{(2)}=2ay^{-1}e^{-2G}*_3d(ye^G).
\end{equation}
The resulting equation is simply $d(y*_3d(\fft1ye^{-G}))=0$, or
\begin{equation}
(\partial_i^2+\ft1y\partial_yy\partial_y){\cal H}=0,
\end{equation}
where ${\cal H}=h^2=\fft1ye^{-G}$ is a function of $(x^1,x^2,y)$.  It is now
evident that ${\cal H}$ is a harmonic function in a four-dimensional
space $\mathbb R^2\times \mathbb R^2$ where $(x^1,x^2)$ span the first
$\mathbb R^2$ and $y$ corresponds to the radial direction in the second
(auxiliary) $\mathbb R^2$.  Since there is no angular dependence in the second
$\mathbb R^2$, the harmonic function is restricted to $s$-wave only solutions
in the auxiliary space.

Putting together the above relations (and taking $a=1$), we find that the
solution has the form
\begin{eqnarray}
ds_6^2&=&-{\cal H}^{-1}(dt^2+d\phi^2)+
{\cal H}(\delta_{ij}dx^idx^j+dy^2+y^2d\tilde\phi^2),\nonumber\\
F_{(2)}&=&2dt\wedge d\fft1{\cal H}.
\label{eq:multicenter}
\end{eqnarray}
Note that, because $dV=0$, we have set $V_i=0$ since this may always be
obtained by a suitable gauge transformation (diffeomorphism).
It is now evident that we have reproduced the familiar multi-centered
string solution in six-dimensions, restricted to singlet configurations
along the $\tilde\phi$ direction, under the assumption that the $S^1$
parameterized by $\phi$ has decompactified.  This configuration arises
naturally from the D1-D5 system with $N_1=N_5$.

\subsubsection{Both $\eta$ and $\widetilde\eta$ vanishing}

Finally, for $n=1$, we could have directly performed a standard Kaluza-Klein
reduction on the circles, which would correspond to setting $\eta=\widetilde
\eta=0$.  Here, the constraint (\ref{eq:abeta}) becomes trivial, so that
$a$ and $b$ may take on arbitrary values.  Assuming at least one of the two
is non-vanishing, then (\ref{eq:ehcon}) implies that $H$ is a constant,
which we take to be zero.  With this choice of $H=0$, we then solve
(\ref{eq:hexp}) for
\begin{equation}
h^{-2}=a^2e^G+b^2e^{-G},
\end{equation}
as well as (\ref{eq:f2form}) for
\begin{equation}
F_{(2)}=-2ae^GdG\wedge(dt+V)+2bh^2*_3dG.
\end{equation}
In addition, (\ref{eq:dv}) gives
\begin{equation}
dV=-2abh^4*_3dG=-\fft1{ab}*_3dz,\qquad z\equiv\fft12\fft{a^2e^G-b^2e^{-G}}
{a^2e^G+b^2e^{-G}},
\end{equation}
provided $ab\ne0$, or simply $dV=0$ otherwise.

For $ab\ne0$, the consistency condition $d^2V=0$ yields a three-dimensional
Laplacian, $d*_3dz=0$, or
\begin{equation}
(\partial_i^2+\partial_y^2)z=0.
\end{equation}
An additional constraint similar to (\ref{eq:bad}), which
arises from the $F_{(2)}$ equation of motion, is still present.  This
time, however it simply states that $dG\wedge *_3dG=0$ in the
three-dimensional Euclidean space, so that $G$ is necessarily a constant.
As a result, we only obtain the Minkowski vacuum in this case.

For, say $b=0$, on the other hand, the above relations reduce to
\begin{equation}
h^{-2}=a^2e^G,\qquad F_{(2)}=-2ade^G\wedge (dt+V),\qquad dV=0.
\end{equation}
The equation of motion for $F_{(2)}$ then gives $d*_3de^{-G}=0$, resulting
in a solution of the form (setting $a=1$)
\begin{eqnarray}
ds_6^2&=&-e^G(dt^2+d\phi^2)
+e^{-G}(\delta_{ij}dx^idx^j+dy^2+d\tilde\phi^2),\nonumber\\
F_{(2)}&=&2dt\wedge d\fft1{e^{-G}}.
\end{eqnarray}
where ${\cal H}=e^{-G}$ is harmonic in $\mathbb R^3$ spanned by $(x^1,x^2,y)$.
This solution is in fact of the same form as (\ref{eq:multicenter}), and
corresponds to a multi-centered string solution.  This time, however, the
Killing symmetry $\partial/\partial\tilde\phi$ is not of an angular type,
and both circles have decompactified.
As a result, we have unfortunately been unable to obtain any new 1/2 BPS
solutions of the minimal ${\cal N}=(1,0)$ system beyond the already familiar
multi-centered string solutions.

\subsection{$S^1\times S^1$ reduction in the presence of a tensor multiplet}

We now explore the possibility of evading the previous
conclusion about the non-existence of a bubbling $AdS_3\times S^3$ solution, 
within the boundary of our ansatz, by enlarging the set of fields, from
minimal 6-dimensional supergravity, to supergravity coupled to a tensor
multiplet. The field content of the tensor multiplet is: dilaton $\Phi$, 
anti-selfdual tensor $H_{\mu\nu\rho}^-$ and dilatino $\lambda$. The dilatino is
Weyl, with opposite chirality than that of the gravitino.

We continue to work with the same metric ansatz as before, while the 
3-form ansatz becomes
\be
H_{(3)}=\Bigl(F_{(2)}+K_{(2)}\Bigr)\wedge d\phi_1 + 
\bigg(\tilde F_{(2)}+\tilde K_{(2)}\bigg)\wedge d\phi_2,
\ee
where $\tilde F_{(2)}= -e^{-G}*_4 F_{(2)}$ and $\tilde K_{(2)}=e^{-G}*_4
K_{(2)}$. Thus $F,\tilde F$ form the self-dual 3-form $H_{(3)}^+$, while
$K,\tilde K$ define the anti-selfdual 3-form $H_{(3)}^-$.
The Bianchi identity and equation of motion read
\be
dH_{(3)}=0,\qquad d\Bigl(e^{\Phi}*_6 H_{(3)}\Bigr)=0.
\ee
Assuming that $\Phi=\Phi(x)$, in terms of the reduced form fields,
these equations become
\bea
\kern-1em&&d(F_{(2)}+K_{(2)})=d(\tilde F_{(2)}+\tilde K_{(2)})=0,\nonumber\\
\kern-1em&&d\Phi\wedge(F_{(2)}-K_{(2)})+d(F_{(2)}-K_{(2)})=
d\Phi\wedge(\tilde F_{(2)}-\tilde K_{(2)})+d(\tilde F_{(2)}-\tilde K_{(2)})=0.
\quad
\label{beom}
\eea
The $\phi_1\phi_2$ component of the Einstein equation yields the constraint 
\be F_{(2)}\wedge F_{(2)}-K_{(2)}\wedge K_{(2)}=0.
\ee

The supersymmetry variations of the supergravity multiplet are the same
as before, (\ref{ssusy}), with the exception that $F,\tilde F$ must be replaced
by $e^{\Phi/2}F,e^{\Phi/2}\tilde F$. Correspondingly, the spinor bilinear
equations (\ref{eq:difdi})--(\ref{eq:difgi})
are modified by means of the same replacement.
The immediate consequence of this observation is that we obtain
as before $f_2=\exp((H+G)/2), f_1=\exp((H-G)/2)$, $K^\mu\partial_\mu$ 
is a Killing vector and $L_\mu dx^\mu$ is still a closed form. Also, the
relation $h^{-2}=f_1^2+ f_2^2$ holds as well. Therefore the metric 
has become once more
$ds^2=h^{-2}(dt+V)^2+h^2(dy^2+\sum_{i,j=1}^2h_{ij} dx^i dx^j)+e^{H+G}d\phi_1^2
+e^{H-G} d\phi_2^2$. We have learned also that
\be
i_K F_{(2)}=-e^{H+G}e^{-\Phi/2}d(H+G), \qquad i_K \tilde F_{(2)}=-e^{H-G}
e^{-\Phi/2} d(H-G)\label{ikf},
\ee
where $i_K$ denotes the inner contraction with the Killing vector 
$K=h^{-2}(dt+V)$,
and 
\be
dK=\frac 12\bigg( e^{H+G} e^{\Phi/2} F_{(2)} + e^{H-G}e^{\Phi/2} \tilde F_{(2)}
\bigg).
\label{keq}
\ee
Substituting $K$ as well as (\ref{ikf}) into the previous equation 
we find the same
differential equation defining $z=\frac 12\tanh(G)$ as (\ref{eq:harmz}).
Thus the metric is identical to the one derived previously in the absence
of the tensor multiplet.

Consistency of (\ref{keq}), namely $d^2 K=0$, combined with Bianchi and
equations of motion (\ref{beom}) leads to  
\be
K_{(2)}\wedge K_{(2)}=0,
\ee
which in turn implies that $F_{(2)}\wedge F_{(2)}=0$. We see that despite
our efforts to avoid the $F_{(2)}\wedge F_{(2)}$ constraint which translates
into the additional non-linear differential equation that $z(x_1,x_2,y)$
had to satisfy, we have to conclude that turning on the tensor multiplet
did not achieve, as one might have hoped, a bubbling $AdS_3\times S^3$
picture.  As mentioned in the introduction, a possible way to evade the
negative conclusion on bubbling AdS$_3\times S^3$ solutions is to allow
for a 4d axion field (arising from $g_{\phi_1\phi_2}$).
In fact, a rather large class of supersymmetric 6d solutions of conical
defect type \cite{Lunin:2002bj} fall into this class of metrics.

\section{Bubbling 1/4 BPS solutions: turning on an axion-dilaton}

In this section we show how the 1/2 BPS family of solutions discovered
recently by Lin, Lunin and Maldacena \cite{Lin:2004nb}  can be modified 
to accommodate a
holomorphic axion-dilaton field. Of course, in doing so we break the amount
of supersymmetry that the new solutions preserve by half.
We will end up with a family of 1/4 BPS solutions which have the same
$SO(4)\times SO(4)$ isometries inherited from the 1/2 BPS family.

Our interest in this class of 1/4 BPS solutions resides in its implications
for the dual gauge theory. We expect that turning on the
axion-dilaton field $\tau$, which amounts to adding D7 branes 
by appropriately including their back-reaction, will lead to
the addition of flavor degrees of freedom to the dual gauge theory.
By embedding $N_f$ D7 branes in the initial $AdS_5\times S^5$ geometry, 
one adds $N_f$ ${\cal N}=2$ hypermultiplets, $Q$, 
in the fundamental of $N_c$ to the ${\cal N}=4$ $SU(N_c)$ dual
gauge theory. The gauge theory superpotential is accordingly modified by
the addition of the hypermultiplets to $\Tr X[Y,Z]+ \bar Q Z Q$.

More precisely, we begin our construction of 1/4 BPS solutions in type
IIB supergravity with the following ansatz:
\bea
ds^2&=&g_{\mu\nu}dx^\mu dx^\nu +e^{H+G}d\Omega_3^2+e^{H-G}d\tilde\Omega_3^2,
\\
F_{(5)}&=&F_{\mu\nu}dx^\mu\wedge dx^\nu\wedge d\Omega_3+\tilde F_{\mu\nu}dx^\mu
\wedge dx^\nu\wedge d\tilde\Omega_3,\\
\tau&=&\tau(x^1+ix^2), \qquad {\rm with}\quad x^\mu=\{t,y,x^1,x^2\}.
\eea
As in \cite{Burrington:2004id}, we will be able to exploit the fact that
the D3-D7 problem separates, with the D7 branes curving the space
transverse to them, and the warping due to the D3 branes modified to
accommodate the D7 branes' backreaction. The self-duality condition and
the Bianchi identity of the 5-form imply for the reduced form fields
\be
F=e^{3G}*_4 \tilde F, \qquad F=dB, \qquad \tilde F=d\tilde B.
\ee
Requiring that this solution is supersymmetric, we impose
\bea
\delta \psi_M&=&(\nabla_M-\frac i2
Q_M)\varepsilon+ \frac{i}{480} F_{M_1 M_2 M_3 M_4 M_5}
\Gamma^{M_1 M_2 M_3 M_4 M_5}{\varepsilon} =0\\
\delta \lambda &=& i P_M \Gamma^M \varepsilon^* =0,
\eea
where $\psi_M$ and $\lambda$ are the complex gravitino and dilatino, whose
$U(1)$ charges are $1/2$ and $3/2$ respectively. The axion and dilaton fields
parameterize a scalar coset $SL(2,R)/U(1)$, with the U(1) connection
given by
\be
Q_M=-\frac 12 \frac{\partial_M\Re\tau}{\Im\tau},
\ee
and where $g^{MN} P_M P_N^*$ represents the kinetic term of the sigma-model 
Lagrangian, with
\be
P_M=-\frac 12\frac{\partial_M \Re\tau}{\Im\tau}.
\ee
Notice that the previous supersymmetry variations of the gravitino along
the sphere directions, (\ref{ch}),~(\ref{cg})
are not modified by the presence of the scalar fields, and that (\ref{psimu})
contains a new term due to the $Q$-connection.
The new constraint following from the supersymmetry variation of the dilatino
only enforces
\be
(\Gamma^1+i\Gamma^2)\varepsilon = (\gamma^1+i\gamma^{2})\epsilon=0
\label{dil}.
\ee
The spinor bilinear equations derived previously (\ref{eq:difdi}) 
are unchanged,
because a bilinear of the type $(\bar\epsilon\ldots\epsilon)$ is U(1) neutral.
However, the one-form
\be
\omega= \epsilon^T C \gamma_\mu \epsilon dx^\mu,
\ee
where $C$ is the charge conjugation matrix ($\gamma^\mu=-C\gamma^{\mu, T}C$),
is no longer closed as it was the case in the absence of the axion-dilaton
field; rather it obeys
\be
d\omega=iQ\wedge\omega\label{om}.
\ee
Given that $K_\mu$ is still a Killing vector and $L_\mu dx^\mu$ is still an
exact form, we can choose as before $K^\mu$ as the generator of time
translations, $K^t=1$, and we choose $L=\gamma dy$, with $\gamma=\pm 1$.
Therefore we arrive at the same form of the metric
\be
ds_4^2 = -h^{-2}(dt+V_i dx^i)^2 + h^2 (dy^2+ \tilde h_{ij} dx^i dx^j).
\ee
We can see now that the constraint (\ref{dil}) has become a projection
condition on $\epsilon$
\be
(1+i\gamma^{\underline{12}})\epsilon=0.
\ee
Using the Killing spinor equations we
find the following set of equations
\bea
f_2 \partial_\mu H&=&-\tilde\eta e^{-(H-G)/2}L_\mu,\\
\partial_\mu f_2 &=& -e^{-3(H+G)/2}F_{\mu\nu}K^\nu=
\frac 12 f_2 \partial_\mu(H+G),
\eea
which allows the integration of both the spinor bilinear $f_2$ and H
as
\be
f_2 = 4\alpha e^{(H+G)/2},\qquad B_t=-\alpha e^{2(H+G)},\qquad e^H=y,
\ee
where we fix the sign of $\gamma$ such that $-\tilde\eta\gamma=1$.
Similarly we find
\be
f_1=4\beta e^{(H-G)/2},\qquad \tilde B_t=-\beta e^{2(H-G)}.
\ee
With the choice $4\beta=1$, we end up by fixing $\alpha$ using on the one hand
(\ref{eq:hexp})
\be
h^{-2}=f_1^2+f_2^2,
\ee
and
\be
\tilde\eta h^{-2}\partial_y e^H = \tilde\eta h^{-2}=\tilde\eta
\bigg(\frac{f_2^2}{4\alpha}-\frac{\eta f_1^2}{4\tilde\eta\beta}\bigg),
\ee
on the other hand. The latter equation is
obtained from (\ref{ch}) by multiplication with $\bar\epsilon\gamma^5$.
We conclude that
the last two equations imply $4\alpha=1$, and  $\eta=-\tilde\eta$.
Substituting $H$ into the Killing spinor equation (\ref{ch}) we identify
another projector
\bea
0&=&\bigg(\frac{1}{yh}\gamma^{\underline3}
+\eta e^{-(H+G)/2}-i\tilde\eta\gamma_5 e^{-(H-G)/2}
\bigg)\epsilon\nonumber\\
&=&\bigg(\sqrt{1+e^{-2G}}\gamma^{\underline3}-\tilde\eta(ie^{-G}
\gamma_5+1)\bigg)\epsilon.
\eea
Moreover, using that
\bea
K^t&=&h\bar\epsilon\gamma^{\underline0}\epsilon=h\epsilon^\dagger\epsilon=1,
\nonumber\\
L_y&=&h\bar\epsilon\gamma^{\underline3}\gamma_5\epsilon=-\tilde\eta,
\eea
one derives the last projector
\be
(\tilde\eta-\gamma^{\underline0}\gamma^{\underline3}\gamma_5)\epsilon=
(\tilde\eta-i\gamma^{\underline{12}})\epsilon=0.
\ee
However, we do not have the freedom of two choices of sign for $\eta$, because
of the first projection condition that we encountered (\ref{dil}) from the
susy variation of the dilatino which identifies
\be
\tilde\eta=-1.
\ee
Thus our solution ends up preserving
only 1/4 of the 32 supersymmetries.

Finally, using the projectors we can solve for the Killing spinor
\be
\epsilon=e^{(H+G)/4}\exp(i\delta\gamma_5\gamma_{\underline3})\epsilon_0,\qquad
\sinh\delta=e^{-G},\qquad \epsilon_0^\dagger\epsilon=1.
\ee
Substituting the Killing spinor in (\ref{om}), where
still the only non-vanishing components are $\omega_{\underline1}$ and
$\omega_{\underline2}$,
we realize that we end up with a conformally flat
two dimensional space in the $x^1, x^2$ directions:
\bea
&&\tilde h_{ij} dx^i dx^j= e^{\Psi(x^1,x^2)} ((dx^1)^2+(dx^2)^2),\\
&&\omega=e^{\Psi(x^1,x^2)}d(x_1+i dx^2).
\eea
Moreover, the conformal factor $e^\Psi$ is related to the axion-dilaton
field, because we argued earlier that the U(1) connection $Q$ becomes
the spin connection in this two dimensional space
\be
\Psi(x_1,x_2)=\Im \tau(x_1+ix_2).
\ee
In fact, there is even more freedom in defining $\Psi(x_1,x_2)$ in terms of 
multiplication by a holomorphic and an antiholomorphic function $\exp(\Psi)=
\Im\tau \exp(f(x_1+ix_2)+f^*(x_1-ix_2))$. This 
stems from the freedom of multiplying $\epsilon_0$ by a phase: $\exp((f-f^*)/2)
\epsilon_0$. Imposing modular invariance, with $N_f$ D7 branes located
at $Z_i=(x_1+ix_2)_i=0$, uniquely determines the conformal factor
as
\be
e^{\Psi(Z,Z^*)}=\Im\tau\bigg|\eta(\tau)\bigg|^4 \prod_{i=1}^{N_f}
\bigg|Z-Z_i\bigg|^{-1/6}.
\ee
The corresponding geometry is non-singular provided that $N_f<24$.  
Near the D7 branes, the axion-dilaton field behaves as a logarithm, 
and its equation of motion has delta-function singularities at the location 
of the D7 branes
\bea
\tau(Z)\approx\frac{i}{g}+\frac{1}{2\pi i}\sum_i\ln(Z-Z_i),\\
\exp(\Psi)\approx\frac{1}{g}-\frac{1}{2\pi}\sum_i\ln(|Z-Z_i|).
\label{imt}
\eea

We are left only with determining the 5-form flux: given the similarity
of our equations to those of \cite{Lin:2004nb}, it is easy now to see that 
the bubbling 1/4 BPS solutions read
\bea
&&e^{-\Psi}\partial_i^2 z+y\partial_y(\frac 1y \partial_y z)=0,\nonumber\\
&&dV=\frac1y *_3 dz, \qquad z=\frac 12\tanh G,\nonumber\\
&&F= dB_t\wedge(dt +V)+ B_t dV+ d\hat B,\qquad
\tilde F= d\tilde B_t\wedge (dt+V)+\tilde B_t dV+ d\hat{\tilde B},\nonumber\\
&&B_t=\frac 14 y^2 e^{2G},\qquad \tilde B_t=-\frac14 y^2 e^{-2G},\nonumber\\
&&d\hat B=\frac{-1}4 y^3 *_3 d\frac{z+1/2}{y^2},\qquad
d\hat{\tilde B}=\frac{-1}4 y^3 *_3 d\frac{z-1/2}{y^2},
\eea
where one should keep in mind that the Hodge symbol $*_3$ in the
three dimensional space is computed relative to the metric
$d y^2 + e^\Psi \sum_{i=1,2} (dx^i)^2$.

To gauge the effect of the axion-dilaton field on the geometry, we can
in a first order of approximation solve the differential 
equation which defines the auxiliary function $z(x_1,x_2,y)$ 
perturbatively in $N_f$. We assume that all D7 branes are overlapping and 
we approximate the conformal factor by $\Im\tau$ (\ref{imt}).
We define polar coordinates in the $(x^1, x^2)$ plane, and we redefine
the radial coordinate $\rho$ by $r=\rho e^{\Psi/2}$. The effect of this 
rescaling is to map the line element $ds_2^2=e^\Psi(d\rho^2+ \rho^2 d\varphi^2)$ into $ds^2\approx(1-\frac{N_f}{2\pi})(dr^2+r^2 d\varphi^2)$. This is 
nothing else but a 2-dimensional space with a deficit angle. Therefore
$z(x_1,x_2,y)/y^2$ is to a first order of approximation still a harmonic 
function, but it is a harmonic function of a 6-dimensional space, with a
deficit angle in the 2-plane defined by $x_1, x_2$. 

Therefore the presence of the D7 branes, while not affecting to a dramatic
degree the bubbling AdS$_5\times S^5$ picture, so that in particular it does 
not change the interpretation of $z(x_1, x_2,y=0)=\pm 1/2$ as a ``phase
space'', nevertheless induces a deficit angle in this plane. Since the
fluctuations
of the D3 branes in the direction $Z=x_1+ix_2$ become in the decoupling limit
the BPS chiral primary operators defining the gauged quantum mechanics 
matrix model of \cite{Corley:2001zk,Berenstein:2004kk}, a deficit angle in the $(x_1, x_2)$
planes translates into a non-trivial monodromy of the chiral 
primary operators. This ultimately implies that the eigenvalues of $Z$ have
a non-trivial monodromy, or equivalently, the electrons participating in 
the quantum Hall effect (i.e. the eigenvalues) have fractional statistics.

\section*{Acknowledgments}
We would like to thank F.~Larsen for many illuminating discussions
throughout the course of this project, as well as A.~Sinkovics
for comments.  We are also grateful to M.~Perry for bringing Weyl
solutions to our attention. This work was supported in part by the
US~Department of Energy under grant DE-FG02-95ER40899.

\appendix
\section{Bosonic reductions on $S^n\times S^n$}
\label{sec:apa}

The reduction of the bosonic fields (metric and form-field) may be performed
in arbitrary dimensions.  For a reduction to four-dimensions on $S^n\times
S^n$, we start with a $D=4+2n$ dimensional bosonic action of the form
\begin{equation}
e^{-1}{\cal L}=R-\fft1{2\cdot (n+2)!}F_{(n+2)}^2.
\label{eq:alag}
\end{equation}
The resulting equations of motion are simply
\begin{eqnarray}
&&R_{MN}=\fft1{2(n+1)!}\left[(F^2)_{MN}-\fft1{2(n+2)}g_{MN}F^2\right],
\nonumber\\
&&dF=d{*F}=0.
\label{eq:aeom}
\end{eqnarray}
Note that here we have taken $F_{(n+2)}$ to be canonically normalized.
Furthermore, at this stage we do not impose self-duality on the form-field,
although below we will show what modifications would be necessary to cover
the self-dual case.

The reduction of the equations of motion, (\ref{eq:aeom}), proceeds by
taking an ansatz of the form
\begin{eqnarray}
&&ds^2=g_{\mu\nu}(x)dx^\mu dx^\nu+e^{H(x)}\left(e^{G(x)}d\Omega_n^2
+e^{-G(x)}d\widetilde\Omega_n^2\right),\nonumber\\
&&F_{(n+2)}=F_{(2)}\wedge\omega_n+\widetilde F_{(2)}(x)\wedge\widetilde
\omega_n,
\label{eq:aans}
\end{eqnarray}
where $\omega_n$ and $\widetilde\omega_n$ are volume forms on the
respective unit $n$-spheres.  Our goal is now to obtain the four-dimensional
effective theory for the fields $g_{\mu\nu}$, $H$, $G$, $F_{(2)}$ and
$\widetilde F_{(2)}$.

At this point, it is worth recalling that, for sphere compactifications,
the fields $H$ and $G$ are actually breathing mode scalars which live in
the massive Kaluza-Klein tower.  In general, it would be inconsistent to
retain only a subset of the massive Kaluza-Klein states, as they typically
source each other {\it ad infinitum}.  However, the scalars $H$ and $G$
themselves are uncharged on the spheres, and hence this breathing mode
reduction remains consistent by virtue of retaining only singlets on the
spheres \cite{Bremer:1998zp,Liu:2000gk}.

We begin by reducing the form-field equation of motion.  From the ansatz
(\ref{eq:aans}), we may obtain the Hodge dual
\begin{equation}
*F_{(n+2)}=*_4e^{nG}\widetilde F_{(2)}\wedge\omega_n+(-)^n*_4e^{-nG}F_{(2)}
\wedge\widetilde\omega_n.
\end{equation}
At this point, we note that, since $*_4*_4=-1$, we may only impose a
self-dual condition on $F_{(n+2)}$ for odd $n$ dimensions ({\it i.e.}~$D=6$
or $10$).  In such dimensions, self-duality yields the relation
$\widetilde F_{(2)}=-*_4e^{-nG}F_{(2)}$.  In any case, we see that the
$F_{(n+2)}$ equation of motion reduces simply to
\begin{eqnarray}
&&dF_{(2)}=0,\qquad d(e^{-nG}*_4F_{(2)})=0,\nonumber\\
&&d\widetilde F_{(2)}=0,\qquad d(e^{nG}*_4\widetilde F_{(2)})=0.
\label{eq:afeom}
\end{eqnarray}

Turning to the Einstein equation, we first compute the Ricci tensor for
the metric ansatz (\ref{eq:aans})
\begin{eqnarray}
R^{(D)}_{\mu\nu}&=&R_{\mu\nu}-\ft{n}2(\partial_\mu H\partial_\nu H
+\partial_\mu G\partial_\nu G)-n\nabla_\mu\nabla_\nu H,\nonumber\\
R^{(D)}_{ab}&=&\hat R_{ab}-\ft{n}2g_{ab}\partial^\mu H\partial_\mu(H+G)
-\ft12g_{ab}\square(H+G),\nonumber\\
R^{(D)}_{\tilde a\tilde b}&=&\hat R_{\tilde a\tilde b}-\ft{n}2g_{\tilde a
\tilde b}\partial^\mu H\partial_\mu(H-G)-\ft12g_{\tilde a\tilde b}
\square(H-G).
\label{eq:a4eins}
\end{eqnarray}
Here $\hat R_{ab}=(n-1)\hat g_{ab}$ and $\hat R_{\tilde a\tilde b}=(n-1)
\hat g_{\tilde a\tilde b}$ are the curvatures on the {\it unit} $n$-spheres
with metrics $\hat g_{ab}$ and $\hat g_{\tilde a\tilde b}$, respectively.
The $D$-dimensional metric components in the sphere directions are
$g_{ab}=e^{H+G}\hat g_{ab}$ and $g_{\tilde a\tilde b}=e^{H-G}\hat g_{\tilde a
\tilde b}$.

For the form-field, we compute
\begin{eqnarray}
F_{(n+2)}^2&=&\ft12(n+2)!\left(e^{-n(H+G)}F^2
+e^{-n(H-G)}\widetilde F^2\right),\nonumber\\
(F_{(n+2)}^2)_{\mu\nu}&=&(n+1)!\left(e^{-n(H+G)}(F^2)_{\mu\nu}+e^{-n(H-G)}
(F^2)_{\mu\nu}\right),\nonumber\\
(F_{(n+2)}^2)_{ab}&=&\ft12(n+1)!e^{-n(H+G)}g_{ab}F^2,\nonumber\\
(F_{(n+2)}^2)_{\tilde a\tilde b}&=&\ft12(n+1)!e^{-n(H-G)}g_{\tilde a\tilde b}
\widetilde F^2,\nonumber\\
(F_{(3)}^2)_{a\tilde a}&=&\hat e_a\hat e_{\tilde a}F_{\mu\nu}\widetilde
F^{\mu\nu}\qquad\hbox{for $n=1$}.
\end{eqnarray}
These expressions allow us to work out the source for the Einstein
equations.  They may be combined with (\ref{eq:a4eins}) to obtain
the four-dimensional equations of motion
\begin{eqnarray}
&&R_{\mu\nu}=\ft{n}2(\partial_\mu H\partial_\nu H+\partial_\mu G\partial_\nu G)
+n\nabla_\mu\nabla_\nu H\nonumber\\
&&\qquad\qquad+\ft12e^{-n(H+G)}(F^2_{\mu\nu}-\ft14g_{\mu\nu}F^2)
+\ft12e^{-n(H-G)}(\widetilde F^2_{\mu\nu}-\ft14g_{\mu\nu}\widetilde F^2),
\nonumber\\
&&\square H+n\partial^\mu H\partial_\mu H=2(n-1)e^{-H}\cosh G,\nonumber\\
&&\square G+n\partial^\mu H\partial_\mu G=-\ft14e^{-n(H+G)}F^2
+\ft14e^{-n(H-G)}\widetilde F^2-2(n-1)e^{-H}\sinh G.
\label{eq:aeeom}
\end{eqnarray}
The scalar equations were separated by taking appropriate linear combinations
of the $R_{ab}$ and $R_{\tilde a\tilde b}$ equations.  In addition, for
$n=1$ ($D=6$), the reduction of the $R_{a\tilde a}$ Einstein equation
yields a constraint $F_{\mu\nu}\widetilde F^{\mu\nu}=0$.  In general, this
signifies an inconsistency in the reduction.  However, so long as we
satisfy this constraint, we are ensured that solutions to the effective
four-dimensional theory may be lifted to solutions of the original
six-dimensional theory.

The equations of motion, (\ref{eq:afeom}) and (\ref{eq:aeeom}), may be
derived from an effective four-dimensional Lagrangian
\begin{eqnarray}
e^{-1}{\cal L}&=&e^{nH}\Bigl[R+\ft12n(2n-1)\partial H^2-\ft12n\partial G^2
-\ft14e^{-n(H+G)}F^2\nonumber\\
&&\qquad-\ft14e^{-n(H-G)}\widetilde F^2+2n(n-1)e^{-H}\cosh G\Bigr].
\label{eq:aslag}
\end{eqnarray}
If desired, this may be transformed into the Einstein frame by the Weyl
transformation $g_{\mu\nu}=e^{-nH}\widetilde g_{\mu\nu}$.  The resulting
Einstein frame action has the form
\begin{eqnarray}
\widetilde e^{-1}{\cal L}&=&\widetilde R-\ft12n(n+1)\partial H^2-\ft12n
\partial G^2-\ft14e^{-nG}F^2-\ft14e^{nG}\widetilde F^2\nonumber\\
&&\qquad+2n(n-1)e^{-(n+1)H}\cosh G.
\label{eq:aelag}
\end{eqnarray}
Note that the scalar fields are not canonically normalized.  Nevertheless,
we find it convenient to retain this convention, so as to avoid unpleasant
factors of $\sqrt{n}$ and $\sqrt{n(n+1)}$.

Finally, for the reductions we have considered, the form field $F_{(n+2)}$
is taken to be self dual in $D=10$ or $6$.  Reducing the self-dual field
follows the procedure given above, so long as we impose the self-dual
condition {\it after} obtaining the equations of motion, (\ref{eq:aeom}). 
In this case, the $F^2$ term vanishes, and we are left with an Einstein
equation of the form
\begin{equation}
R_{MN}=\fft1{2(n+1)!}(F^2)_{MN}.
\label{eq:sdeom}
\end{equation}
Note that, if canonical normalization is desired, we ought to include an
additional factor of $1/2$ in the field-strength term of the original
Lagrangian, (\ref{eq:alag}), in which case the right-hand side of
(\ref{eq:sdeom}) must also be multiplied by $1/2$.  This is indeed what
we do for the IIB theory.  However, we forego this factor of $1/2$ for
the ${\cal N}=(1,0)$ model in six dimensions.  This choice of a
non-canonically normalized 3-form $H_{(3)}$ avoids $\sqrt{2}$ factors in
the supersymmetry transformation (\ref{eq:6gto}) of section~\ref{sec:s1s1}
(and furthermore keeps canonical normalization in the case where the
${\cal N}=(1,0)$ theory is coupled to a single tensor multiplet).

Regardless of normalization, for the self-dual case, we impose the
condition $\widetilde F_{(2)}=-*_4e^{-nG}F_{(2)}$ to eliminate
$\widetilde F_{(2)}$ in favor of $F_{(2)}$ in the reduced
equations of motion.  For canonical self-dual normalization, which
incorporates the additional factor of $1/2$ introduced above, this
simply amounts to erasing all $\widetilde F$ terms from the expressions
in (\ref{eq:aeeom}).  The resulting effective Lagrangians are identical
to (\ref{eq:aslag}) and (\ref{eq:aelag}), except with the $\widetilde F^2$
terms removed.  If we instead leave out the factor of $1/2$, the resulting
$F^2$ terms are twice as large (and the $\widetilde F^2$ terms are
absent as usual).  Here, we see the familiar feature that while
$\widetilde F$ cannot be dualized in the Lagrangian, it is valid to do
so in the equations of motion.

In addition, for self-duality in $D=6$,
the constraint $F_{\mu\nu}\widetilde F^{\mu\nu}=0$ is replaced by
$F_{(2)}\wedge F_{(2)}=0$.  Here it is clear that $F_{(2)}\wedge F_{(2)}$
would ordinarily source
an axionic field.  However, by truncating away all axions, we can no
longer allow such a source.  Again, so long as we impose this constraint
by hand, we will still be able to obtain solutions to the original
six-dimensional model.

\section{Integrability of the Killing spinor equations}
\label{sec:apb}

Since we have found the somewhat surprising result that the Killing spinor
equations resulting from both $S^3\times S^3$ compactification of IIB
supergravity and $S^1\times S^1$ compactification of the ${\cal N}=(1,0)$
theory have very similar forms, it is interesting to see how they can
lead to different equations of motion, namely (\ref{eq:aeeom}) with either
$n=3$ or $n=1$.  In order to see how this works, we may consider the
integrability of the Killing spinor equations, (\ref{eq:4kse}),
repeated here as
\begin{equation}
\delta\psi_\mu={\cal D}_\mu\epsilon,\qquad
\delta\chi_H=\Delta_H,\qquad\delta\chi_G=\Delta_G,
\end{equation}
where
\begin{eqnarray}
{\cal D}_\mu&=&\nabla_\mu+\ft{i}{16}e^{-\fft{n}2(H+G)}F_{\nu\lambda}
\gamma^{\nu\lambda}\gamma_\mu,\nonumber\\
\Delta_H&=&\gamma^\mu\partial_\mu H+e^{-\fft12H}(\eta e^{-\fft12G}
-i\widetilde\eta\gamma_5e^{\fft12G}),\nonumber\\
\Delta_G&=&\gamma^\mu\partial_\mu G
+e^{-\fft12H}(\eta e^{-\fft12G}+i\widetilde\eta\gamma_5e^{\fft12G})
-\ft{i}4e^{-\fft{n}2(H+G)}F_{\mu\nu}\gamma^{\mu\nu}\epsilon.
\label{eq:ddsusy}
\end{eqnarray}

In the original theory (either in $D=10$ or $6$), there is only one object
to examine for first order integrability, namely $[{\cal D}_M,{\cal D}_N]$.
However, viewed in the effective four-dimensional point of view, we may
consider the various commutators of ${\cal D}_\mu$, $\Delta_H$ and
$\Delta_G$.  We begin with $[{\cal D}_\mu,{\cal D}_\nu]$.  After
straightforward although tedious manipulations, we find
\begin{eqnarray}
\gamma^\mu[{\cal D}_\mu,{\cal D}_\nu]&=&
\ft12[R_{\nu\sigma}-\ft{n+1}8e^{-n(H+G)}(F^2{}_{\nu\sigma}-\ft14g_{\mu\nu}F^2)
\nonumber\\
&&\qquad
-\ft{n}2(\partial_\nu H\partial_\sigma H+\partial_\nu G\partial_\sigma G)
-n\nabla_\nu\nabla_\sigma H]\gamma^\sigma\nonumber\\
&&+\ft{i}{16}e^{-\fft{n}2(H+G)}\partial_{[\mu}F_{\lambda\sigma]}
\gamma^{\mu\lambda\sigma}\gamma_\nu+\ft{i}8e^{-\fft{n}2(H-G)}
\nabla^\mu(e^{-nG}F_{\mu\lambda})\gamma^\lambda\gamma_\nu\nonumber\\
&&+\ft{n}2[{\cal D}_\nu,\Delta_H]+\ft{n}4\partial_\nu H\Delta_H
+\ft{n}4\partial_\nu G\Delta_G\nonumber\\
&&-\ft{in}{32}e^{-\fft{n}2(H+G)}F_{\lambda\sigma}
\gamma^{\lambda\sigma}\gamma_\nu(\Delta_H-\Delta_G).
\label{eq:caldd}
\end{eqnarray}
Since the last two lines above vanish on Killing spinors, we see that
this integrability yields the Einstein equation as well as Bianchi and
equation of motion for $F_{(2)}$.  In particular, if the latter two
conditions are imposed on $F_{(2)}$, then the Einstein equation is
guaranteed by supersymmetry.  Note also that the Einstein equation of
(\ref{eq:aeeom}) is reproduced with proper dimension dependent ($n=1$
or $3$) coefficients.  This also shows the curious fact that, starting
from an identical normalization of $F_{(2)}$ in the supersymmetry
variations, (\ref{eq:ddsusy}), one in fact obtains different normalizations
in the bosonic equations involving $F_{(2)}$.

Turning to the $[{\cal D}_\mu,\Delta_H]$ condition, we find
\begin{eqnarray}
\gamma^\mu[{\cal D}_\mu,\Delta_H]&=&\square H+n\partial H^2-(n-1)e^{-H}
(\eta^2e^{-G}+\widetilde\eta^2e^G)\nonumber\\
&&+[-n\gamma^\mu\partial_\mu H-\ft{i}8e^{-\fft{n}2(H+G)}F_{\mu\nu}
\gamma^{\mu\nu}\nonumber\\
&&\qquad+(n-\ft12)e^{-\fft12H}(\eta e^{-\fft12G}+i\widetilde\eta
\gamma_5e^{\fft12G})]\Delta_H\nonumber\\
&&-\ft12e^{-\fft12H}(\eta e^{-\fft12G}-i\widetilde\eta\gamma_5e^{\fft12G})
\Delta_G,
\label{eq:caldh}
\end{eqnarray}
which simply reproduces the $H$ equation of motion in (\ref{eq:aeeom}).
In particular, for $n\ne1$ we are required to choose $\eta^2=\widetilde\eta^2
=1$, while for $n=1$ these $U(1)$ charges are irrelevant.  This
demonstrates that the identical bosonic equations are satisfied, regardless
of the Kaluza-Klein charges carried by the Killing spinors.

At this stage, it is also worth noting that we may form the combination
\begin{equation}
[\gamma^\mu\partial_\mu H-e^{-\fft12H}(\eta e^{-\fft12G}+i\widetilde\eta
\gamma_5e^{\fft12G})]\Delta_H
=\partial H^2-e^{-H}(\eta^2e^{-G}+\widetilde\eta^2e^G).
\end{equation}
When acting on Killing spinors, this demonstrates that supersymmetry
further imposes the condition
\begin{equation}
\partial H^2=e^{-H}(\eta^2e^{-G}+\widetilde\eta^2e^{G}).
\end{equation}
Combining this with the equation of motion of $H$ yields the simple
expression
\begin{equation}
\square H+\partial H^2=0,
\label{eq:hcons}
\end{equation}
which must be satisfied on supersymmetric backgrounds.

The final integrability condition we obtain is the one between ${\cal D}_\mu$
and $\Delta_G$.  In this case, we obtain
\begin{eqnarray}
\gamma^\mu[{\cal D}_\mu,\Delta_G]&=&\square G+n\partial H\partial G
+\ft{n+1}{16}e^{-n(H+G)}F^2-(n-1)e^{-H}(\eta^2e^{-G}-\widetilde\eta^2
e^{G})\nonumber\\
&&\qquad +\ft{3-n}{32}e^{-n(H+G)}F_{\mu\nu}F_{\lambda\sigma}
\gamma^{\mu\nu\lambda\sigma}\nonumber\\
&&-\ft{i}4e^{-\fft{n}2(H+G)}\partial_{[\mu}F_{\nu\lambda]}
\gamma^{\mu\nu\lambda}-\ft{i}2e^{-\fft{n}2(H-G)}\nabla^\mu(e^{-nG}F_{\mu\nu})
\gamma^\nu\nonumber\\
&&+\ft12[-n\gamma^\mu\partial_\mu G+(n-1)e^{-\fft12H}(\eta e^{-\fft12G}
-i\widetilde\eta\gamma_5e^{\fft12G})]\Delta_H\nonumber\\
&&+\ft12[-n\gamma^\mu\partial_\mu H+(n-1)e^{-\fft12H}(\eta e^{-\fft12G}
+i\widetilde\eta\gamma_5e^{\fft12G})\nonumber\\
&&\qquad+\fft{i(n-1)}4e^{-\fft{n}2(H+G)}
F_{\mu\nu}\gamma^{\mu\nu}]\Delta_G.
\label{eq:caldg}
\end{eqnarray}
In addition to the $G$ equation as well as Bianchi and equation
of motion for $F_{(2)}$, we see that the $F_{(2)}\wedge F_{(2)}=0$ constraint
shows up in this integrability condition provided $n\ne3$.  So, at
least for the ${\cal N}=(1,0)$ theory, supersymmetry implies not just
the equations of motion of (\ref{eq:aeeom}), but also the $F_{(2)}\wedge
F_{(2)}=0$ constraint.

More precisely, for partially broken supersymmetry, the Killing spinor
equations often yield only linear combinations of the equations of
motion.  In this case, we see from (\ref{eq:caldh}) that both the
$H$ equation as well as the $H$ constraint (\ref{eq:hcons}) are automatically
satisfied independent of the rest of the fields.  However, from
(\ref{eq:caldd}) we see that the Einstein equation is only satisfied in
combination with the $F_{(2)}$ equations, and similarly from (\ref{eq:caldg}),
that the $G$ equation of motion is satisfied in combination with the
$F_{(2)}$ equations.
We may conclude that, for obtaining supersymmetric backgrounds, it would
be sufficient to satisfy the $F_{(2)}$ Bianchi identity and equation of
motion in addition to the Killing spinor equations themselves.

Finally, while the supersymmetry transformations (\ref{eq:4kse}) were only
obtained for the cases $n=3$ and $1$, they may nevertheless be formally
extended to any value of $n$.  Examination of the integrability conditions
(\ref{eq:caldd}), (\ref{eq:caldh}) and (\ref{eq:caldg}) indicate
consistency with a bosonic sector described by an effective Lagrangian
\begin{eqnarray}
e^{-1}{\cal L}&=&e^{nH}\Bigl[R+\ft12n(2n-1)\partial H^2-\ft12n\partial G^2
-\ft{n+1}{16}e^{-n(H+G)}F^2\nonumber\\
&&\qquad+n(n-1)e^{-H}(\eta^2e^{-G}+\widetilde\eta^2e^G)\Bigr].
\end{eqnarray}
For $n\ne3$ this system must be extended with the constraint $F_{(2)}
\wedge F_{(2)}=0$.

\section{Differential identities for the spinor bilinears}
\label{sec:apc}

The supersymmetric construction of
\cite{Gauntlett:2002sc,Gauntlett:2002nw,GMR,Gauntlett:2004zh} proceeds
by postulating the existence of a Killing spinor $\epsilon$ and then
forming the tensors $f_1$, $f_2$, $K_\mu$, $L_\mu$ and $Y_{\mu\nu}$ from
spinor bilinears (\ref{eq:bilinear}).  The algebraic identities of interest
were given in the text in (\ref{eq:h2}).  Here we tabulate the differential
identities obtained by demanding that $\epsilon$ solves the Killing
spinor equations (\ref{eq:4kse}).

First, by assuming $\delta\psi_\mu=0$, we may demonstrate that
\begin{eqnarray}
\nabla_\mu f_1&=&\ft14e^{-\fft{n}{2}(H+G)}*F_{\mu\nu}K^\nu,\nonumber\\
\nabla_\mu f_2&=&-\ft14e^{-\fft{n}{2}(H+G)}F_{\mu\nu}K^\nu,\nonumber\\
\nabla_\mu K_\nu&=&\ft14e^{-\fft{n}{2}(H+G)}(f_2F_{\mu\nu}-f_1*F_{\mu\nu}),
\nonumber\\
\nabla_\mu L_\nu&=&\ft14e^{-\fft{n}{2}(H+G)}
(\ft12g_{\mu\nu}F_{\lambda\rho}Y^{\lambda\rho}-2F_{(\mu}{}^\lambda
Y_{\nu)\lambda}),\nonumber\\
\nabla_\mu Y_{\nu\lambda}&=&\ft14e^{-\fft{n}2(H+G)}(2g_{\mu[\nu}
F_{\lambda]\rho}L^\rho-2F_{\mu[\nu}L_{\lambda]}+F_{\nu\lambda}L_\mu).
\label{eq:difdi}
\end{eqnarray}
In particular, the equation for $K_\mu$ indicates that $K_{(\mu;\nu)}=0$,
so that $K^\mu$ is Killing.  This is in fact a generic feature of
constructing a Killing vector from Killing spinors.

In addition, the $\delta\chi_H=0$ condition allows us to derive the
additional relations
\begin{eqnarray}
&&K^\mu\partial_\mu H=0,\kern9.3em
\eta f_2=-\widetilde\eta e^Gf_1,\nonumber\\
&&L^\mu\partial_\mu H=\eta e^{-\fft12(H+G)}f_1-\widetilde\eta
e^{-\fft12(H-G)}f_2,\nonumber\\
&&\eta e^{-\fft12(H+G)}L_\mu=f_1\partial_\mu H,\kern4.3em
\widetilde\eta e^{-\fft12(H-G)}L_\mu=-f_2\partial_\mu H,\nonumber\\
&&\eta e^{-\fft12(H+G)}K_\mu=*Y_\mu{}^\nu\partial_\nu H,\kern3em
\widetilde\eta e^{-\fft12(H-G)}K_\mu=Y_\mu{}^\nu\partial_\nu H,\nonumber\\
&&2L_{[\mu}\partial_{\nu]}H=0,\kern3em
2K_{[\mu}\partial_{\nu]}H=\eta e^{-\fft12(H+G)}*Y_{\mu\nu}
+\widetilde\eta e^{-\fft12(H-G)}Y_{\mu\nu}.
\label{eq:difhi}
\end{eqnarray}
Similarly, the $\delta\chi_G=0$ condition yields the relations
\begin{eqnarray}
&&K^\mu\partial_\mu G=0,\kern10em
\ft14e^{\fft{1-n}2(H+G)}F_{\mu\nu}*Y^{\mu\nu}=\eta f_2-\widetilde\eta e^Gf_1,
\nonumber\\
&&L^\mu\partial_\mu G=\eta e^{-\fft12(H+G)}f_1+\widetilde\eta e^{-\fft12(H-G)}
f_2-\ft14e^{-\fft{n}2(H+G)}F_{\mu\nu}Y^{\mu\nu},\nonumber\\
&&\eta e^{-\fft12(H+G)}L_\mu=f_1\partial_\mu G+\ft12e^{-\fft{n}2(H+G)}
*F_{\mu\nu}K^\nu,\nonumber\\
&&\kern4em
\widetilde\eta e^{-\fft12(H-G)}L_\mu=f_2\partial_\mu G+\ft12e^{-\fft{n}2(H+G)}
F_{\mu\nu}K^\nu,\nonumber\\
&&\eta e^{-\fft12(H+G)}K_\mu=*Y_\mu{}^\nu\partial_\nu G+\ft12e^{-\fft{n}2(H+G)}
*F_{\mu\nu}L^\nu,\nonumber\\
&&\kern4em
\widetilde\eta e^{-\fft12(H-G)}K_\mu=-Y_\mu{}^\nu\partial_\nu G
+\ft12e^{-\fft{n}2(H+G)}F_{\mu\nu}L^\nu,\nonumber\\
&&2L_{[\mu}\partial_{\nu]}G=2e^{-\fft{n}2(H+G)}F_{[\mu}{}^\rho Y_{\nu]\rho},
\nonumber\\
&&2K_{[\mu}\partial_{\nu]}G=\eta e^{-\fft12(H+G)}*Y_{\mu\nu}
-\widetilde\eta e^{-\fft12(H-G)}Y_{\mu\nu}-\ft12e^{-\fft{n}2(H+G)}
(f_1*F_{\mu\nu}+f_2F_{\mu\nu}).\qquad
\label{eq:difgi}
\end{eqnarray}
Although the above identities are algebraic and not differential
on the spinor bilinears, they originate from the supersymmetry variations
along the internal directions of the Kaluza-Klein reduction.  So in this
sense, they form a generalized set of `differential identities'.  However,
as they are only algebraic, they prove extremely useful in determining
much of the geometry, as is evident from the analysis of \cite{Lin:2004nb}.


\end{document}